\documentclass[conference]{IEEEtran}
\IEEEoverridecommandlockouts
\usepackage{amsmath,amssymb,amsfonts}
\usepackage{algorithmic}
\usepackage{graphicx}
\usepackage{textcomp}
\usepackage{xcolor}

\usepackage{amsmath,amssymb,amsfonts}
\usepackage{algorithmic}
\usepackage{graphicx}
\usepackage{textcomp}
\usepackage{xcolor}

\usepackage{amsmath}

\newcommand\matlab{MATLAB\textregistered}
\usepackage[capitalise, noabbrev]{cleveref}
\usepackage[utf8]{inputenc}

\usepackage{glossaries}
\makenoidxglossaries%

\newacronym{2d}{2D}{two-dimensional}
\newacronym{3d}{3D}{three-dimensional}
\newacronym{3gpp}{3GPP}{3rd generation partnership project}
\newacronym{4g}{4G}{fourth-generation}
\newacronym{5g}{5G}{fifth-generation}
\newacronym{6g}{6G}{sixth-generation}
\newacronym{abp}{ABP}{authentication by personalisation}
\newacronym{ac}{AC}{alternating current}
\newacronym{aclr}{ACLR}{adjacent channel leakage ratio}
\newacronym{adc}{ADC}{analog-to-digital converter}
\newacronym{adr}{ADR}{adaptive data rate}
\newacronym{afe}{AFE}{analog front end}
\newacronym{afhd2a}{AFHDS2A}{Automatic Frequency Hopping Digital System Second Generation}
\newacronym{ag}{AG}{array gain}
\newacronym{ai}{AI}{artificial intelligence}
\newacronym{am}{AM}{amplitude modulation}
\newacronym{amam}{AM/AM}{amplitude modulation to amplitude modulation}
\newacronym{amp}{AMP}{approximate message passing}
\newacronym{ampm}{AM/PM}{amplitude modulation to phase modulation}
\newacronym{aoa}{AOA}{angle-of-arrival}
\newacronym{aod}{AOD}{angle-of-departure}
\newacronym{ap}{AP}{access point}
\newacronym{apu}{APU}{access point unit}
\newacronym{ar}{AR}{augmented reality}
\newacronym{arp}{ARP}{Antenna Reference Point}
\newacronym{asic}{ASIC}{application specific integrated circuit}
\newacronym{ask}{ASK}{amplitude-shift keying}
\newacronym{auv}{AUV}{autonomous underwater vehicle}
\newacronym{awgn}{AWGN}{additive white Gaussian noise}
\newacronym{baw}{BAW}{bulk acoustic wave}
\newacronym{bb}{BB}{base-band}
\newacronym{bc}{BC}{backscatter communication}
\newacronym{bd}{BD}{backscatter device}
\newacronym{ber}{BER}{bit error rate}
\newacronym{bf}{BF}{beamforming}
\newacronym{bldc}{BLDC}{brushless DC}
\newacronym{ble}{BLE}{Bluetooth Low Energy}
\newacronym{bom}{BOM}{bill of materials}
\newacronym{bs}{BS}{base station}
\newacronym{bw}{BW}{bandwidth}
\newacronym{ca}{CA}{carrier emitter}
\newacronym{cad}{CAD}{channel activity detection}
\newacronym{cars}{CARS}{calibration reference signal}
\newacronym{cbm}{CBM}{condition based maintenance}
\newacronym{cc}{CC}{constant current}
\newacronym{ccdf}{CCDF}{complementary cumulative distribution function}
\newacronym{ccnn}{CCNN}{circular convolutional neural network}
\newacronym{ccs}{CCS}{correlative channel sounder}
\newacronym{cdf}{CDF}{cumulative distribution function}
\newacronym{cdrx}{CDRX}{connected mode DRX}
\newacronym{ce}{CE}{coverage enhancement}
\newacronym{ced}{CED}{cumulative energy density}
\newacronym{cf}{CF}{cell-free}
\newacronym{cfmmimo}{CF-mMIMO}{cell-free massive MIMO}
\newacronym{cfo}{CFO}{carrier frequency offset}
\newacronym{cir}{CIR}{channel impulse response}
\newacronym{cmos}{CMOS}{complementary metal oxide semiconductor}
\newacronym{cost}{COST}{commercial off-the-shelf}
\newacronym{cots}{COTS}{commercial off-the-shelf}
\newacronym{cp}{CP}{cyclic prefix}
\newacronym{cpt}{CPT}{capacitive power transfer}
\newacronym{cpu}{CPU}{central-processing unit}
\newacronym{cqi}{CQI}{channel quality indicator}
\newacronym{cr}{CR}{coding rate}
\newacronym{crc}{CRC}{cyclic redundancy check}
\newacronym{crlb}{CRLB}{Cram\'er-Rao lower bound}
\newacronym{crs}{CRS}{cell reference signal}
\newacronym{cs}{CS}{compressed sensing}
\newacronym{cse}{CSE}{channel state estimation}
\newacronym{csi}{CSI}{channel state information}
\newacronym{csp}{CSP}{contact service point}
\newacronym{css}{CSS}{chirp spread spectrum}
\newacronym{cu}{CU}{central unit}
\newacronym{cv}{CV}{constant voltage}
\newacronym{cw}{CW}{continuous wave}
\newacronym{dac}{DAC}{digital-to-analog converter}
\newacronym{daq}{DAQ}{data acquisition system}
\newacronym{das}{DAS}{distributed antenna systems}
\newacronym{dc}{DC}{direct current}
\newacronym{dcc}{DCC}{dynamic cooperation clustering}
\newacronym{de}{DE}{drain efficiency}
\newacronym{dl}{DL}{downlink}
\newacronym{dlc}{DLC}{Distributed Laser Charging}
\newacronym{dli}{DLI}{direct link interference}
\newacronym{dm}{DM}{diffuse multipath}
\newacronym{dma}{DMA}{Direct Memory Access}
\newacronym{dmc}{DMC}{diffuse multipath component}
\newacronym{dmimo}{D-MIMO}{distributed MIMO}
\newacronym{doa}{DOA}{direction-of-arrival}
\newacronym{dpd}{DPD}{digital pre-distortion}
\newacronym{dpdk}{DPDK}{Data Plane Development Kit}
\newacronym{drx}{DRX}{Discontinuous Reception Mode}
\newacronym{duc}{DUC}{digital up-converter}
\newacronym{e2e}{E2E}{end-to-end}
\newacronym{easa}{EASA}{European Union Aviation Safety Agency}
\newacronym{ec}{EC}{European Commission}
\newacronym{ecdf}{eCDF}{empirical cumulative distribution function}
\newacronym{ecsp}{ECSP}{edge computing service point}
\newacronym{edlc}{EDLC}{electrostatic double-layer capacitors}
\newacronym{edrx}{eDRX}{Extended Discontinuous Reception Mode}
\newacronym{ee}{EE}{energy efficiency}
\newacronym{egprs}{EGPRS}{Enhanced Data Rates for GSM Evolution}
\newacronym{eh}{EH}{energy harvesting}
\newacronym{eirp}{EIRP}{equivalent isotropically radiated power}
\newacronym{em}{EM}{electromagnetic}
\newacronym{embb}{eMBB}{enhanced Mobile Broadband}
\newacronym{en}{EN}{energy neutral}
\newacronym{end}{END}{energy neutral device}
\newacronym{eol}{EoL}{end of life}
\newacronym{epu}{EPU}{edge processing unit}
\newacronym{erp}{ERP}{equivalent radiated power}
\newacronym[plural=ESCs,firstplural=electronic speed controllers (ESCs)]{esc}{ESC}{electronic speed control}
\newacronym{esd}{ESD}{electrostatic discharge}
\newacronym{etsi}{ETSI}{European Telecommunications Standards Institute}
\newacronym{evd}{EVD}{eigenvalue decomposition}
\newacronym{evm}{EVM}{Error Vector Magnitude}
\newacronym{fa}{FA}{federation anchor}
\newacronym{fdma}{FDMA}{frequency division multiple access}
\newacronym{fem}{FEM}{finite element}
\newacronym{fet}{FET}{field-effect transistor}
\newacronym{fembb}{feMBB}{further enhanced mobile broadband}
\newacronym{fft}{FFT}{fast Fourier transform}
\newacronym{fh}{FH}{fronthaul}
\newacronym{fim}{FIM}{Fisher information matrix}
\newacronym{fom}{FoM}{figure of merit}
\newacronym{fpga}{FPGA}{field-programmable gate array}
\newacronym{fsk}{FSK}{frequency shift keying}
\newacronym{fss}{FSS}{frequency selective surface}
\newacronym{gan}{GaN}{galliumnitride}
\newacronym{gb}{GB}{grant-based}
\newacronym{gf}{GF}{grant-free}
\newacronym{gnb}{gNB}{Next Generation Node B}
\newacronym{gnn}{GNN}{graph neural network}
\newacronym{gnss}{GNSS}{global navigation satellite system}
\newacronym{gpclk}{GPCLK}{general purpose clock}
\newacronym{gprs}{GPRS}{General Packet Radio Services}
\newacronym{gps}{GPS}{Global Positioning System}
\newacronym{gpu}{GPU}{graphical processing unit}
\newacronym{gscm}{GSCM}{geometry‐based stochastic model}
\newacronym{gsm}{GSM}{Global System for Mobile Communications}  %
\newacronym{gwp}{GWP}{Global Warming Potential}
\newacronym{harq}{HARQ}{hybrid automatic repeat request}
\newacronym{hat}{HAT}{hardware attached on top}
\newacronym{hcs}{HCS}{human-centric services}
\newacronym{hpbm}{HPBM}{half power beam width}
\newacronym{HyMPRo}{HyMPRo}{Hybrid Multi-Path Routing algorithm}
\newacronym{i.i.d.}{i.i.d.}{independent and identically distributed}
\newacronym{ib}{IB}{in-band}
\newacronym{ibo}{IBO}{input back-off}
\newacronym{ic}{IC}{integrated circuit}
\newacronym{id}{ID}{information decoding}
\newacronym{if}{IF}{intermediate-frequency}
\newacronym{iid}{i.i.d.}{independently and identically distributed}
\newacronym{iis}{IIS}{integrated information system}
\newacronym{im}{IM}{intermodulation}
\newacronym{imu}{IMU}{inertial measurement unit}
\newacronym{io}{IO}{input/output}
\newacronym{iot}{IoT}{Internet of Things}
\newacronym{ipt}{IPT}{inductive power transfer}
\newacronym{ipy}{IPY}{Interventions per Year}
\newacronym{iq}{IQ}{in-phase and quadrature}
\newacronym{ir}{IR}{infrared}
\newacronym{ism}{ISM}{industrial, scientific and medical}
\newacronym{isp}{ISP}{internet service provider}
\newacronym{jfet}{JFET}{junction field effect transistor}
\newacronym{kpi}{KPI}{key performance indicator}
\newacronym{kvi}{KVI}{key value indicator}
\newacronym{lca}{LCA}{life cycle assessment}
\newacronym{lco}{LCO}{lithium cobalt oxide}
\newacronym{ldo}{LDO}{Low-dropout voltage regulator}
\newacronym{ldpc}{LDPC}{low-density parity-check}
\newacronym{less}{LESS}{Low Energy Scheduler Solution}
\newacronym{lfp}{LFP}{lithium iron phosphate}
\newacronym{lic}{LIC}{lithium-ion capacitor}
\newacronym{lidar}{LiDAR}{light detection and ranging}
\newacronym{liion}{Li-ion}{lithium-ion}
\newacronym{lipo}{LiPo}{lithium polymer}
\newacronym{lis}{LIS}{large intelligent surface}
\newacronym{llh}{LLH}{log-likelihood}
\newacronym{lmmse}{LMMSE}{least minimum mean square error}
\newacronym{lmo}{LMO}{lithium ion manganese oxide}
\newacronym{lo}{LO}{local oscillator}
\newacronym{lora}{LoRa}{long range}
\newacronym{lorawan}{LoRaWAN}{long-range wide-area network}
\newacronym{los}{LoS}{line-of-sight}
\newacronym{lp}{LP}{linear programming}
\newacronym{lpt}{LPT}{laser power transfer}
\newacronym{lpwa}{LPWA}{Low Power Wide Area}
\newacronym{lpwan}{LPWAN}{low-power wide-area network}
\newacronym{lqi}{LQI}{link quality indicator}
\newacronym{lrelu}{LReLU}{leaky rectified linear unit}
\newacronym{lrt}{LRT}{likelihood-ratio test}
\newacronym{ls}{LS}{least squares}
\newacronym{lsa}{LSA}{large synthetic array}
\newacronym{lsf}{LSF}{large-scale fading}
\newacronym{lsfc}{LSFC}{large-scale fading component}
\newacronym{ltc}{LTC}{lithium thionyl chloride}
\newacronym{lte}{LTE}{Long Term Evolution}
\newacronym{lti}{LTI}{linear time-invariant}
\newacronym{lto}{LTO}{lithium titanate}
\newacronym{m2m}{M2M}{machine to machine}
\newacronym{mac}{MAC}{Medium Access Control}
\newacronym{mate}{MATE}{millimeter-wave MIMO testbed}
\newacronym{mc}{MC}{Monte Carlo}
\newacronym{mcl}{MCL}{Maximum Coupling Loss}
\newacronym{mcs}{MCS}{modulation and coding scheme}
\newacronym{mcu}{MCU}{Microcontroller Unit}
\newacronym{mec}{MEC}{multi-access edge computing}
\newacronym{mems}{MEMS}{micro-electromechanical systems}
\newacronym{mf}{MF}{matched filter}
\newacronym{mimo}{MIMO}{multiple-input multiple-output}
\newacronym{miso}{MISO}{multiple-input single-output}
\newacronym{ml}{ML}{machine learning}
\newacronym{mlp}{MLP}{multilayer perceptron}
\newacronym{mmimo}{mMIMO}{massive MIMO}
\newacronym{mmse}{MMSE}{minimum mean square error}
\newacronym{mmtc}{mMTC}{massive machine-typed communication}
\newacronym{mmwave}{mmWave}{millimeter wave}
\newacronym{mn}{MN}{matching network}
\newacronym{mosfet}{MOSFET}{metal-oxide semiconductor field effect transistor}
\newacronym{mpc}{MPC}{multipath component}
\newacronym{mppt}{MPPT}{maximum power point tracking}
\newacronym{mr}{MR}{maximum ratio}
\newacronym{mrc}{MRC}{magnetic resonance coupling}
\newacronym{mrt}{MRT}{maximum ratio transmission}
\newacronym{mse}{MSE}{mean square error}
\newacronym{mtc}{MTC}{Machine-Type Communication}
\newacronym{multi-rat}{Multi-RAT}{multiple radio access technology}
\newacronym{music}{MUSIC}{MUltiple SIgnal Classification}
\newacronym{nb}{NB}{narrowband}
\newacronym{nbiot}{NB-IoT}{narrowband IoT}
\newacronym{nca}{NCA}{nickel cobalt aluminum}
\newacronym{nfc}{NFC}{near-field communication}
\newacronym{nicd}{NiCd}{nikkel cadmium}
\newacronym{nimh}{NiMH}{nikkel metal hydride}
\newacronym{nlos}{NLoS}{non-line-of-sight}
\newacronym{nmc}{NMC}{nickel manganese cobalt}
\newacronym{nmos}{nMOS}{n-channel metal-oxide semiconductor}
\newacronym{nn}{NN}{neural network}
\newacronym{nnls}{NNLS}{non-negative least squares}
\newacronym{noma}{NOMA}{non-orthogonal multiple access}
\newacronym{np}{NP}{Neyman-Pearson}
\newacronym{npbch}{NPBCH}{Narrowband Physical Broadcast Channel}
\newacronym{npss}{NPSS}{Narrow Band Primay Synchronization Signal}
\newacronym{nr}{NR}{New Radio}
\newacronym{nrs}{NRS}{Narrow Band Reference Signal}
\newacronym{ntp}{NTP}{network time protocol}
\newacronym{oai}{OAI}{OpenAirInterface} 
\newacronym{ofdm}{OFDM}{orthogonal frequency-division multiplexing}
\newacronym{ofdmim}{OFDM-IM}{OFDM with index modulation}
\newacronym{oob}{OOB}{out-of-band}
\newacronym{ook}{OOK}{on-off keying}
\newacronym{oran}{O-RAN}{open radio-access network}
\newacronym{os}{OS}{operating system}
\newacronym{ota}{OTA}{over-the-air}
\newacronym{otaa}{OTAA}{over-the-air authentication}
\newacronym{p1}{P1}{Phase 1}
\newacronym{p2}{P2}{Phase 2}
\newacronym{p2p}{P2P}{point-to-point}
\newacronym{pa}{PA}{power amplifier}
\newacronym{pae}{PAE}{power-added efficiency}
\newacronym{pana}{PanA}{Panel A}
\newacronym{panb}{PanB}{Panel B}
\newacronym{papr}{PAPR}{peak-to-average power ratio}
\newacronym{pc}{PC}{pilot count}
\newacronym{pcb}{PCB}{printed circuit board}
\newacronym{pcg}{PCG}{power consumption gain}
\newacronym{pcsi}{PCSI}{perfect channel state information}
\newacronym{pd}{PD}{powered device}
\newacronym{pdcch}{PDCCH}{physical downlink control channel}
\newacronym{pdf}{PDF}{probability density function}
\newacronym{pdp}{PDP}{power delay profile}
\newacronym{pdsch}{PDSCH}{physical downlink shared channel}
\newacronym{per}{PER}{packet error rate}
\newacronym{pg}{PG}{path gain}
\newacronym{pgd}{PGD}{proximal gradient descent}
\newacronym{pl}{PL}{path loss}
\newacronym{pll}{PLL}{phase-locked loop}
\newacronym{poe}{PoE}{power-over-Ethernet}
\newacronym{pps}{1PPS}{pulse per second}
\newacronym{prbs}{PRBs}{Physical Resource Blocks}
\newacronym{ps}{PS}{Processing System}
\newacronym{psd}{PSD}{power spectral density}
\newacronym{pse}{PSE}{power sourcing equipment}
\newacronym{psk}{PSK}{phase shift keying}
\newacronym{psm}{PSM}{power saving mode}
\newacronym{pss}{PSS}{primary synchronisation signal}
\newacronym{ptp}{PTP}{precision-time protocol}
\newacronym{ptrs}{PTRS}{Phase-Tracking Reference Signals}
\newacronym{ptw}{PTW}{paging time window}
\newacronym{ptu}{PTU}{power transmitting unit}
\newacronym{pru}{PRU}{power receiving unit}
\newacronym{pw}{PW}{planar wavefront}
\newacronym{pwm}{PWM}{pulse width modulation}
\newacronym{qam}{QAM}{quadrature amplitude modulation}
\newacronym{qos}{QoS}{quality-of-service}
\newacronym{quadriga}{QuaDRiGa}{QUAsi Deterministic RadIo channel GenerAtor}
\newacronym{ra}{RA}{Random Access}
\newacronym{ran}{RAN}{radio access network}
\newacronym{rar}{RAR}{Random Access Response}
\newacronym{rat}{RAT}{radio access technology}
\newacronym{rbs}{RBS}{radio base station}
\newacronym{rcs}{RCS}{radar cross section}
\newacronym{re}{RE}{radio element}
\newacronym{relu}{ReLU}{rectified linear unit}
\newacronym{rf}{RF}{radio frequency}
\newacronym{rfeh}{RFEH}{radio frequency energy harvesting}
\newacronym{rfic}{RFIC}{radio-frequency integrated circuit}
\newacronym{rfid}{RFID}{radio frequency identification}
\newacronym{rfpt}{RFPT}{radio frequency power transfer}
\newacronym{rfsoc}{RFSoC}{Radio Frequency System-on-Chip}
\newacronym{ris}{RIS}{reflective intelligent surface}
\newacronym{rllmtc}{RLLMTC}{reliable low latency machine type communication}
\newacronym{rms}{RMS}{root mean square}
\newacronym{rmse}{RMSE}{root-mean-square error}
\newacronym{rmt}{RMT}{random matrix theory}
\newacronym{rof}{RoF}{radio-over-fiber}
\newacronym{ros}{ROS}{robot operating system}
\newacronym{rpi}{RPi}{Raspberry Pi}
\newacronym{rrc}{RRC}{Radio Resource Connection}
\newacronym{rreq}{RREQ}{route request packet}
\newacronym{rsrp}{RSRP}{Reference Signals Received Power}
\newacronym{rss}{RSS}{received signal strength}
\newacronym{rssi}{RSSI}{received signal strength indicator}
\newacronym{rtc}{RTC}{real time clock}
\newacronym{rtk}{RTK}{real time kinematics}
\newacronym{rts}{RTS}{ray tracing simulator}
\newacronym{rv}{RV}{random variable}
\newacronym{rw}{RW}{RadioWeaves}
\newacronym{rx}{RX}{receiver}
\newacronym{rzf}{RZF}{regularized zero forcing}
\newacronym{s-parameter}{S-parameter}{scattering parameter}
\newacronym{sa}{SA}{synchronization anchor}
\newacronym{sar}{SAR}{specific absorption rate}
\newacronym{scfdma}{SCFDMA}{single-carrier frequency division multiple access}
\newacronym{scpi}{SCPI}{Standard Commands for Programmable Instruments}
\newacronym{sdg}{SDG}{Sustainable Development Goal}
\newacronym{sdm}{SDM}{sigma-delta modulator}
\newacronym{sdn}{SDN}{software-defined network}
\newacronym{sdof}{SDoF}{sigma-delta over fiber}
\newacronym{sdr}{SDR}{software-defined radio}
\newacronym{sepic}{SEPIC}{single-ended primary-inductor converter}
\newacronym{sf}{SF}{spreading factor}
\newacronym{sfn}{SFN}{single frequency network}
\newacronym{sfp}{SFP}{small form-factor pluggable}
\newacronym{simo}{SIMO}{single-input multiple-output}
\newacronym{sinr}{SINR}{signal-to-interference-plus-noise ratio}
\newacronym{siso}{SISO}{single-input single-output}
\newacronym{slam}{SLAM}{simultaneous localization and mapping}
\newacronym{slc}{SLC}{spatial leakage suppression}
\newacronym{smc}{SMC}{specular multipath component}
\newacronym{smps}{SMPS}{switched mode power supply}
\newacronym{sndr}{SNDR}{signal-to-noise-and-distortion ratio}
\newacronym{snidr}{SNIDR}{signal-to-noise-and-interference-and-distortion ratio}
\newacronym{snr}{SNR}{signal-to-noise ratio}
\newacronym{soc}{SoC}{state of charge}
\newacronym{sysoc}{SoC}{system-on-chip}
\newacronym{srd}{SRD}{short-range device}
\newacronym{ssb}{SSB}{synchronisation signal block}
\newacronym{ssd}{SSD}{solid state drive}
\newacronym{steam}{STEAM}{science, technology, engineering, the arts, and mathematics}
\newacronym{svd}{SVD}{singular value decomposition}
\newacronym{sw}{SW}{spherical wavefront}
\newacronym{swipt}{SWIPT}{simultaneous wireless information and power transfer}
\newacronym{tau}{TAU}{tracking area update}
\newacronym{tcer}{TCER}{transported to consumed energy ratio}
\newacronym{tdd}{TDD}{time division duplexing}
\newacronym{tdoa}{TDOA}{time-difference-of-arrival}
\newacronym{toa}{TOA}{time-of-arrival}
\newacronym{tof}{ToF}{time-of-flight}
\newacronym{tosm}{TOSM}{through-open-short-match}
\newacronym{trl}{TRL}{technology readyness level}
\newacronym{trp}{TRP}{Transmission Reception Point}
\newacronym{tsn}{TSN}{time-sensitive networking}
\newacronym{ttm}{TTM}{time to market}
\newacronym{ttn}{TTN}{The Things Network}
\newacronym{tx}{TX}{transmitter}
\newacronym{uav}{UAV}{unmanned aerial vehicle}
\newacronym{udp}{UDP}{User Datagram Protocol}
\newacronym{ue}{UE}{user equipment}
\newacronym{ugv}{UGV}{unmanned ground vehicle}
\newacronym{uhd}{UHD}{USRP hardware driver}
\newacronym{uhf}{UHF}{ultra-high frequency}
\newacronym{ul}{UL}{uplink}
\newacronym{ula}{ULA}{uniform linear array}
\newacronym{ummtc}{umMTC}{ultra massive machine type communication}
\newacronym{un}{UN}{United Nations}
\newacronym{upa}{UPA}{uniform planar array}
\newacronym{ura}{URA}{uniform rectangular array}
\newacronym{urllc}{URLLC}{ultra-reliable low-latency communications}
\newacronym{usrp}{USRP}{universal software radio peripheral}
\newacronym{uv}{UV}{unmanned vehicle}
\newacronym{uwb}{UWB}{ultrawideband}
\newacronym{v2v}{V2V}{vehicle-to-vehicle}
\newacronym{vep}{VEP}{virtual edge platform}
\newacronym{vlc}{VLC}{visible light communication}
\newacronym{vlp}{VLP}{visible light positioning}
\newacronym{vna}{VNA}{vector network analyzer}
\newacronym{vr}{VR}{virtual reality}
\newacronym{wb}{WB}{wideband}
\newacronym{wban}{WBAN}{wireless body area network}
\newacronym{wpt}{WPT}{wireless power transfer}
\newacronym{wr}{WR}{White Rabbit}
\newacronym{wrsn}{WRSN}{wireless rechargeable sensor network}
\newacronym{wsn}{WSN}{wireless sensor network}
\newacronym{xets}{XETS}{cross exponentially tapered slot}
\newacronym{xr}{XR}{extended reality}
\newacronym{z3ro}{Z3RO}{zero third-order distortion}
\newacronym{zf}{ZF}{zero-forcing}
\newacronym{zmcscg}{ZMCSCG}{zero mean circularly symmetric complex Gaussian}
\newacronym{zvs}{ZVS}{zero voltage switching}

\usepackage[per-mode=symbol]{siunitx}  
\DeclareSIUnit{\dBm}{dBm}	                
\DeclareSIUnit{\dBi}{dBi}                   
\DeclareSIUnit{\dBsm}{dBsm}                 

\usepackage{tikz}
\usepackage{pgfplots}
\usepackage{circuitikz}
\usetikzlibrary{backgrounds}
\usetikzlibrary{calc}
\usepackage{pgf-pie}

\usepackage{caption}
\usepackage{subcaption}

\usepackage{array, booktabs, xltabular}

\usepackage[backend=biber, style=ieee, citestyle=numeric-comp, maxcitenames=1,mincitenames=1, maxbibnames=1, minbibnames=1,isbn=false, doi=false,citecounter=true]{biblatex}
\AtBeginBibliography{\footnotesize}

\newcommand{\jona}[1]{{\color{orange}[JONA: #1]}}

\usepackage{siunitx}
\DeclareSIUnit \amperehour {Ah} 
\DeclareSIUnit\year{year}
\DeclareSIUnit\years{years}
\DeclareSIUnit\month{month}
\DeclareSIUnit\months{months}

\begin{document}

\title{UAV-Based Solution for Extending the Lifetime of IoT Devices: Efficiency, Design and Sustainability
}

\author{
    \IEEEauthorblockN{Jarne Van Mulders\IEEEauthorrefmark{1}, Sam Boeckx\IEEEauthorrefmark{1}, Jona Cappelle\IEEEauthorrefmark{1}, Liesbet Van der Perre\IEEEauthorrefmark{1} and Lieven de Strycker\IEEEauthorrefmark{1}}
    \IEEEauthorblockA{\IEEEauthorrefmark{1}\textit{KU Leuven,} Belgium, name.surname@kuleuven.be}
}


\maketitle

\begin{abstract}
\acrfull{iot} technology is named as a key ingredient in the evolution towards digitization of many applications and services. A deployment based on battery-powered remote \acrfull{iot} devices enables easy installation and operation, yet the autonomy of these devices poses a crucial challenge. A too short lifespan is undesirable from a functional, economical, and ecological point of view. This paper presents a \acrfull{uav}-based approach to recharge remote \acrfull{iot} nodes. An in-depth study of the charging efficiency and optimization of key parameters, and measurements-based verification, is reported on. An actual corresponding design and implementation of the full \acrshort{uav}-based charging system and its proof-of-concept validation are presented. Finally, the sustainability of the proposed solution is discussed. The results presented in this paper hence confirm that the proposed \acrshort{uav}-based approach and design are functionally successful and efficient charging can be achieved, provided the constraints and challenges coming with the approach are adequately dealt with.  Moreover, it comes with an overall reduction in ecological footprint for \acrshort{iot} applications relying on battery-powered nodes in need of medium energy and/or considerable lifetime expectation (5 years or more).
\end{abstract}

\begin{IEEEkeywords}
Internet of Things, Wireless power transmission, Energy efficiency, Carbon footprint, Life cycle assessment
\end{IEEEkeywords}

\section{Introduction}
\label{section:introduction}
\subsection{Research context}
The deployment of \gls{iot} systems and \glspl{wsn} opens up an interesting potential for environmental monitoring, improving logistics efficiency, and diverse other applications~\cite{Salam2020}. Many \gls{iot} devices thereto need to operate on a battery in a remote location. This causes a main challenge in the adoption of \gls{iot} technology: the nodes have a limited lifetime~\cite{callebaut2021art}, which possibly incurs a high overhead in maintenance. Moreover, the ecological footprint may be considerable as non-rechargeable batteries are often selected because of their high energy density in terms of energy per unit volume and/or weight. 

 The conventional strategies to prolong the autonomy of \gls{iot} devices are: 1) select physically larger batteries, 2) equip the device with an energy harvesting solutions and a rechargeable battery~\cite{SHAIKH20161041,HarvestOverv2018}, which increases the complexity, and 3) plan for service visits to replace or recharge batteries. It should be noted that these three options all come at a cost, while they may not be a physically convenient option for all applications, for example it may be cumbersome to reach \gls{iot} nodes on high positions.

 \begin{figure}
    \centering
    \includegraphics[width=0.5\textwidth]{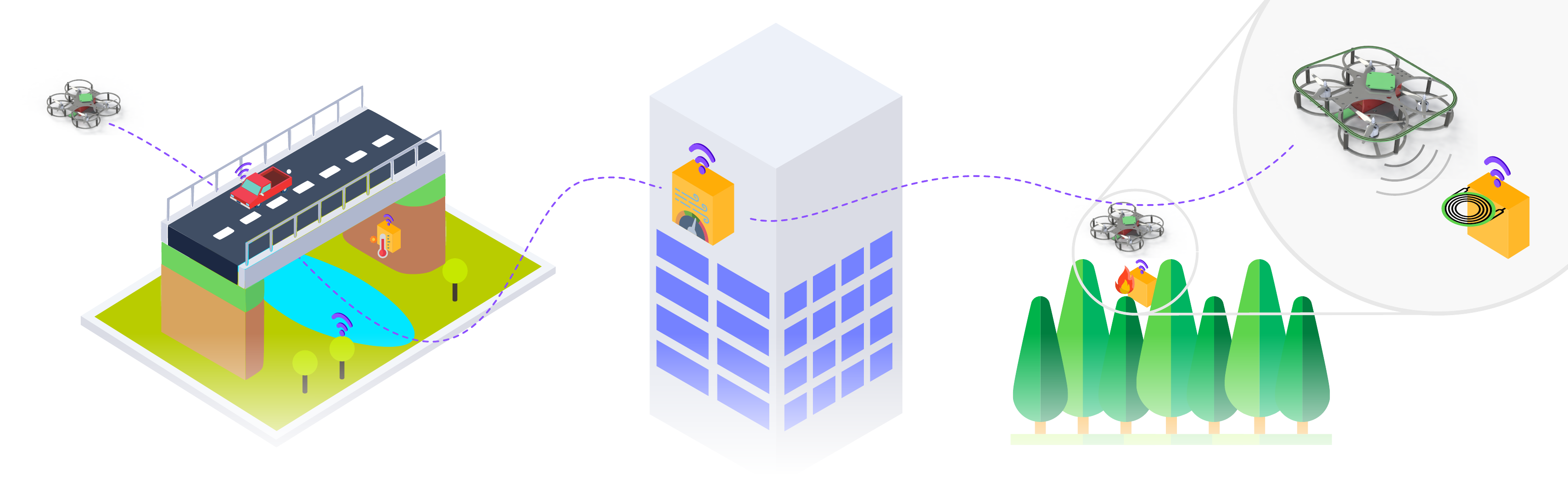}
    \caption{System overview: a UAV approach to extend the lifetime of IoT devices in different use-cases.}
    \label{fig:system_overview}
\end{figure}
 
We here present an \gls{uav}-based charging solution, illustrated in Figure~\ref{fig:system_overview}, as an alternative for extending the lifetime of \gls{iot} devices. 

\subsection{Prior art and related work}

The support of \glspl{uav} in \gls{iot} systems has been proposed mostly to provide network functionality~\cite{HENTATI2020103451}, e.g., for time-varying and vehicle networks~\cite{HEMMATI2023200226} \cite{Cui_TimeVary}
or to perform remote monitoring tasks by themselves~\cite{ASADZADEH2022109633}~\cite{9217835}. In the context of charging, the \gls{uav} is mostly considered as the device to be (wirelessly) charged~\cite{rohan2019advanced}. When \Glspl{uav} are considered in the role of charge provider~\cite{Cetinkaya}~\cite{ojha2023wireless} it most often concerns RF-based charging or RF energy harvesting~\cite{yao2019energy}, whether or not combined with communication, e.g., in \gls{swipt}, which is different from the here considered coupled wireless power transfer. 

While many works have studied the ground-to-\gls{uav} channel for wireless communication purposes over at least several meters (e;g., \cite{Jiang3D}, these models do not apply to the approach discussed in this paper. In this work, we envision that the \gls{uav} approaches the \gls{iot} device closely --typically to within \SI{10}{\centi\meter}. The conceptual idea of a UAV to recharge an \gls{iot} device at its location `on the spot' or replace its battery has been introduced in \cite{van2023uav}, where an efficiency comparison of a coupled wireless power transfer versus RF-based charging is also presented.
By calculating the energy efficiency for this method, we demonstrate that the `\gls{uav} flying in' to charge the device provides an effective option to extend the autonomy of devices. In this paper we investigate the question: To which extent can the charging by the \gls{uav} nearing the device result in autonomy improvements and create an efficient solution? We thereby extend the analysis beyond efficiency of the energy transfer, and also assess the self-impact of the \gls{uav} and the \gls{iot} node to get a more complete picture of the actual environmental impact. By extending the autonomy, we inherently avoid generating new devices or having to replace the battery. At the same time, this approach enables new possibilities for devices with a higher energy consumption in remote areas, where human intervention comes with significant overhead. We here consider the scenario where the \gls{uav} in a dedicated mission charges one \gls{iot} device. We note that further efficiency gains could be achieved when deploying multiple \glspl{uav} and optimizing routes, as for the communications case discussed in~\cite{Khel_routing}.

\subsection{Paper contributions}
 
The focus of this paper is on an in-situ charging solution for \gls{iot} devices accomplished by a \gls{uav} that flies close enough to enable coupled wireless power transfer. The latter can achieve a transfer efficiency on the link that is order of magnitudes higher than RF-waves based transfer. The main contributions of our study are: We detail the charging concept, identify the main additional technological challenges that come into play when targeting a \gls{uav}-based charging solution, and propose a design to optimize specifically the coupled wireless power transfer link. We thereto perform an analytical study of the coupling factor, and focus on the magnetic resonance coupling approach that we identify as most adequate for the problem at hand. We propose a proof of concept prototype, that is compliant with the 250~gram weight limit for unlicensed operation of the \gls{uav} in Europe, according to the regulatory framework by the \gls{easa}~\cite{easa_drone}, and that is able to transfer power in the order of Watts to the \gls{iot} device. The experimental validation of the \gls{uav}-based coupled wireless charging concept is presented. We further provide a critical view of the system implementation by assessing the sustainability aspects.

The remainder of this paper is structured as follows. The next section elaborates on the charging concept and architecture, and the analytical study of the wireless power transfer. The design of the \gls{uav}, capable of \gls{wpt} is presented in section~\ref{section:design}, and the experimental validation in section~\ref{section:measurements}. The analysis of sustainability aspects follows in section~\ref{section:sustainability-analysis}, and the conclusions are formulated in section~\ref{section:conclusion}. 

\section{UAV based charging approach and high-level architecture}
\label{section:arch}

The concept of \gls{uav}-based charging as depicted in \cref{fig:system_overview}, shows that the \gls{uav} is expected to fly to the depleted node and recharge the internal battery. In an autonomous implementation, a system to align the \gls{uav} and the \gls{iot} node is required. While in conventional systems charging connectors are the preferred solution due to high power transfer levels and efficiency, they are susceptible to corrosion, moisture ingress, and the tolerance for alignment errors is quite limited. Selecting a wireless power transfer solution instead, can isolate the electronics from ambient influences and increase the alignment margins, and is therefore considered the most suitable solution for the charging approach handled in this paper. The alignment of the \gls{uav} towards the \gls{iot} device itself is not addressed in this study. We direct the reader to \cite{ojha2023wireless,van2022aerial} for an overview of both coarse and precise positioning as well as alignment capabilities.

The required alignment accuracy depends closely on the chosen \gls{wpt} technology. A comprehensive analysis of potential \gls{wpt} solutions is beyond the scope of this manuscript, and the reader is referred to \cite{van2022wireless} for more information. We can briefly summarize that due to limitations in power density, a coupled system is preferred over an uncoupled system~\cite{van2023uav}. Additionally, a \gls{wpt} system based on coupled capacitive plates is generally only feasible for bridging a few millimeters, making it unsuitable for \gls{uav} implementations. In contrast, \gls{wpt} using fluctuating magnetic fields can cover ranges of upto a few centimeters and is more resilient against misalignments. Therefore, the latter is considered the most viable solution in this context.

\begin{figure}[h]
    \centering
    \resizebox{0.5\textwidth}{!}{

\begin{tikzpicture}[american voltages, every text node part/.style={align=center}]

\ctikzset{bipoles/capacitor/height=0.4}
\ctikzset{bipoles/capacitor/width=0.1}

\tikzset{block/.style = {rectangle, draw=black!50, fill=black!5, thick, minimum width=2cm, minimum height = 3cm}}
\tikzset{medium/.style = {rectangle, densely dotted, draw=black!50, thick, minimum width=2.5cm, minimum height = 4.2cm}}

\ctikzset{resistor = american}




\tikzset{block/.style = {rectangle, draw=black!50, fill=black!5, thick, minimum width=3cm, minimum height = 1.5cm}}
\tikzset{medium/.style = {rectangle, densely dotted, draw=black!50, thick, minimum width=2.5cm, minimum height = 4.2cm}}
\tikzset{small/.style = {rectangle, draw=black!50, fill=black!5, thick, minimum width=2cm, minimum height = 1cm}}
\tikzset{small1/.style = {rectangle, densely dotted, draw=black!50, thick, minimum width=2cm, minimum height = 1cm}}
\tikzset{small2/.style = {rectangle, densely dotted, draw=black!50, thick, minimum width=1.5cm, minimum height = 0.75cm}}
\tikzset{large/.style = {rectangle, dashed, draw=black!50, thick, minimum width=3.1cm, minimum height=4cm}}

\draw[densely dotted] (1.5,0.5)-- +(0,0.3) -- +(0.2625,0.65) -- +(-0.2625,0.65) -- +(0,0.3);
\draw[blue!40] (3.5,0.5)-- +(0,0.3) -- +(0.2625,0.65) -- +(-0.2625,0.65) -- +(0,0.3);
\draw[blue!40] (9.5,0.5)-- +(0,0.3) -- +(0.2625,0.65) -- +(-0.2625,0.65) -- +(0,0.3);
\draw[dotted] (11.5,0.5)-- +(0,0.3) -- +(0.2625,0.65) -- +(-0.2625,0.65) -- +(0,0.3);
\draw[densely dotted] (0.5,2.5)-- +(0,0.3) -- +(0.2625,0.65) -- +(-0.2625,0.65) -- +(0,0.3);
\draw[blue!40] (3.5,2.5)-- +(0,0.3) -- +(0.2625,0.65) -- +(-0.2625,0.65) -- +(0,0.3);
\draw[dotted] (12.5,2.375)-- +(0,0.3) -- +(0.2625,0.65) -- +(-0.2625,0.65) -- +(0,0.3);

\draw (1,0.2) -- (0.3,0.2) to [V] (0.3,-2.2) -- (1,-2.2);
\draw (1,0.2) -- (3.5,0.2);
\draw (1,-2.2) -- (3.5,-2.2);
\draw (4.5,0.2) to[C, l=$C_{1}$] (5.5,0.2) to[L, l_=$L_{1}$] (5.5,-2.2) -- (4.5,-2.2);
\draw (8.5,-2.2) -- (7.5,-2.2) to[L, l_=$L_{2}$] (7.5,0.2) to[C, l=$C_{2}$] (8.5,0.2);
\draw (10,0.2) -- (10.5,0.2);
\draw (10,-2.2) -- (10.5,-2.2);

\draw []
(4.8,-3) -- (4.8,-3.2) -- (8.2,-3.2) -- (8.2,-3)
(2.75,-3.5) -- (2.75,-3.7) -- (12.25,-3.7) -- (12.25,-3.5)
(6,-1) edge[Triangle-Triangle, bend left] node [above] {$k$} (7,-1)
;

\node [block,rotate=90] at (1.75,-1) () {UAV\\ Components};
\node [block,rotate=90] at (3.75,-1) () {WPT\\ Transmitter};
\node [block,rotate=90] at (9.25,-1) () {WPT\\ Receiver};
\node [block,rotate=90] at (11.25,-1) () {IoT\\ Circuit};

\node [small1] at (1,2) () {Flight\\ Controller};
\node [small] at (4,2) () {Data\\ Logger};
\node [small2] at (12,2) () {Gateway};

\node[large] at (4.35,-0.7) () {};
\node[large] at (8.65,-0.7) () {};

\node[align=left] at (6.5,-3) () {Section 3};
\node[align=left] at (5.3,-2.45) () {\tiny PTU};
\node[align=left] at (7.7,-2.45) () {\tiny PRU};
\node[align=left] at (7.5,-3.5) () {Section 4};



\end{tikzpicture}%
}
    \caption{Focused structure of the contributions of this paper. The data logger, \acrshort{wpt} transmitter, and receiver use \acrshort{ble} to communicate. The selected \acrshort{ble} \acrshort{sysoc} can also be employed as application \acrshort{mcu} for the \acrshort{iot} circuit.
    }
    \label{fig:high-level-overview}
\end{figure}
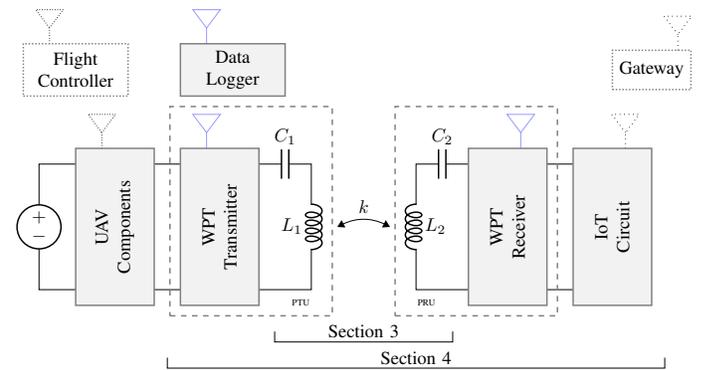

Consequently, the system studied in this paper is limited to a \gls{uav} with a \gls{wpt} transmitter, a receiver and transmitter coil, a \gls{wpt} receiver, and the \gls{iot} node itself. This high-level architecture is presented in \cref{fig:high-level-overview}. The \gls{wpt} transmitter and \gls{wpt} receiver, or \gls{ptu} and \gls{pru} respectively, can communicate wirelessly with each other and optionally send data to a `data logger' for debug and design purposes. For experiments, the `flight controller' is utilized to send flight commands to the \gls{uav}. Lastly, the `gateway' in the \gls{wsn} acts as the \gls{lpwan}-to-internet bridge, centralizing and storing the \gls{iot} data to, e.g., a database. The gateway and flight controller are not further discussed in this paper, as this work focuses on the \gls{uav}-based charging concept.

As mentioned in \cref{section:introduction}, the feasibility of the system will be demonstrated with a \gls{uav} that has weight constraints. Hence, we consider applications for which the energy requirements of the \gls{iot} device are relatively small. A \SI{60}{\milli\amperehour} \gls{lto} cell \cite{LTO_battery_datasheet} is selected to power the \gls{iot}'s onboard sensors, actuators, and \gls{lpwan} modem. Assuming that the \gls{uav} hovers during the charging cycle, it is recommended to limit the charge time. This clarifies the choice of an \gls{lto} cell, which can be recharged at rates of up to \SI{10}{C} (\SI{6}{\minute}), whereas many other cell technologies are typically rechargeable around \SI{1}{C} (\SI{1}{\hour}).

The remainder of this section focuses on the wireless power transfer technology.  As motivated above, \gls{wpt} based on fluctuating magnetic fields is the preferred solution in this context. The achievable power transfer levels and efficiencies depend on the design of the coils and their relative position. The coupling factor between the transmit and receive coil is a central determining factor, including its sensitivity to alignment errors.

Two technologies, namely \gls{ipt} and \gls{mrc}, provide energy transmission via magnetic fields. The most suitable solution depends on the feasible coupling factors between the transmit and receive coils. In \ref{subsection:coupling-prediction}, an analytical study is performed to estimate the coupling factors. 

\subsection{Coupling factor analytical study}
\label{subsection:coupling-prediction}

To optimize the coil selection and parameter settings, an analytical study is performed to predict the coupling factor based on simplified coil models and \matlab{} simulations. The resulting estimated coupling factors can provide initial insight into the feasibility of the system and time-consuming \gls{fem} analyses can be avoided in the first place. In \cref{section:coil-design}, advanced simulation and measurement techniques are applied as a verification and to obtain more accurate coupling factor results.

This analysis is confined to a set of assumptions pertaining to the specific parameters of the transmit and receive coils.

\begin{itemize}
    \item The dimensions of the \textbf{transmit coil} are constrained by the size of the \gls{uav} model, which measure $146$\,×\,\SI{164}{\milli\meter}.
    \item Given the spatial constraints of the \gls{uav} frame, it is feasible to incorporate \textbf{two windings} on the \textbf{transmit side}, without altering the flight characteristics of the \gls{uav}.
    \item The quantity of \textbf{windings} on the \textbf{receiver} side may exceed that on the transmitter side. However, an excessive number of windings can impact the quality factor due to increased equivalent resistance. The quality factor depends on the frequency and affects the link performance, as discussed further in \cref{subsection:mrc-intro}. Moreover higher self-inductance values potentially result in more tuning instabilities due to reduced tuning capacitor values.
    \item Given the assumption that the \gls{uav} hovers during the charging process, it is crucial to ensure that the receiver coil does \textbf{not obstruct the airflow} completely.
\end{itemize}

\subsubsection{Mathematical model for coupling coefficient estimations}
\label{subsubsection:math-k}

The analytical models, based on formulas from \cite{rosa} and \cite{Liu2019}, which we implemented in \matlab{} for simulation purposes, provide a baseline indication of the coupling between two coils. To make the simulation less time-consuming than \gls{fem} analyses, spiral coils are approximated by concentric circles.

The coupling coefficient $k$ between two coils is calculated as follows
\begin{equation}
k = \frac{M_{12}}{\sqrt{L_1 \cdot L_2}}
\label{eqn:k}
\end{equation}
with $M_{12}$ the mutual inductance between the two coils, $L_1$ the self-inductance of the primary coil and $L_2$ the self-inductance of the secondary coil.

To calculate the self-inductance $L$ of a coil, the self-inductance $L_i$ of every winding needs to be calculated as well as the mutual inductance $M_{ij}$ between every possible winding pair. The sum off these, results in the total self-inductance of the coil. 
For a coil with two windings $i$ and $j$ for instance, the self-inductance results in $L = L_i + L_j + M_{ij} + M_{ji}$. Note that $M_{ij}$ and $M_{ji}$ are equal.

Equation \eqref{eq:L1} is used to calculate the self-inductance of every winding \cite{rosa}.
\begin{equation}
    \label{eq:L1}
    L_i = \mu r_i \left(\ln\left(\frac{8r_i}{a}\right)-2+\frac{1}{4}Y\right) 
\end{equation}
with $\mu$ the magnetic permeability, equal to $\mu_0 \mu_r$ where $\mu_0$ is the magnetic permeability in vacuum with a value of $4 \pi \cdot 10^{-7}$ H/m and $\mu_r$ is the relative magnetic permeability of the medium which is air in this case, so, $\mu_r$ is here equal to one. $r_i$ is the radius of the winding and $a$ is the radius of the wire. The term $Y$ accounts for the skin effect.
\begin{equation}
    Y= \frac{1}{1+a\sqrt{\frac{1}{8}\mu\sigma\omega}}
\end{equation}
with $\omega = 2 \pi f$ the angular frequency and $\sigma$ the specific conductivity of the wire.
$Y$ will be zero if the current only flows on the surface of the wire and one if the current is distributed uniformly over the cross section of the wire.

Equation \eqref{eqn:M} is used to calculate the mutual inductance between a pair of windings~\cite{Liu2019}
\begin{equation}
\label{eqn:M}
    M_{ij} = \mu_0 \ \sqrt{r_ir_j} \left[\left(\frac{2}{s}-s\right)K(s)-\frac{2}{s}E(s)\right]
\end{equation}
with
\begin{equation*}
\label{eqn:s}
 s = \sqrt{\frac{4r_ir_j}{(r_i+r_j)^2+d^2}}   
\end{equation*}
and $r_i$ the radius of one winding, $r_j$ the radius of another winding, $d$ the vertical distance between the centers of the wires and $K(s)$ and $E(s)$ the complete elliptic integrals of the first and second kind respectively
\begin{equation}
\label{eqn:ellipse}
    K(s) = \int_{0}^{\frac{\pi}{2}}\frac{1}{\sqrt{1-s^2\sin^2\theta}}d\theta
\end{equation}
and
\begin{equation*}
    E(s) = \int_{0}^{\frac{\pi}{2}}{\sqrt{1-s^2\sin^2\theta}}d\theta.
\end{equation*}
In this case the vertical distance $d = 0$, seeing that the coils exist of only one layer of windings. The windings are all in the same plane.

The mutual inductance $M_{12}$ between two coils can be calculated in the same way. Equation \eqref{eqn:M} is used again yet $r_i$ and $r_j$ are radii of windings of different coils and $d$ is now the vertical distance between the two coils. This is calculated for every possible winding pair and summed to get the overall mutual inductance. For two coils with two windings for instance, the mutual inductance results in $M_{12} = M_{1i2i} + M_{1j2j} + M_{1i2j} + M_{1j2i}$.

Knowing $L_1$, $L_2$ and $M_{12}$ makes it possible to calculate $k$ using \eqref{eqn:k}.

\subsubsection{Calculating coupling factors for the considered use case}

The \matlab{}-based study is conducted at a frequency of \SI{6.78}{\mega\hertz}, a license-free \gls{ism} band. In fact, the self-inductance and coupling factors are mainly influenced by the coil specifications and the distance between the transmitter and receiver, with a lesser impact from the frequency. In contrast, the skin effect, depending on the frequency, affects the equivalent series resistance of the coil and consequently also the quality factor. This latter is important for the link efficiency, discussed in \cref{subsection:wpt-link-eff}. The skin effect has a minor impact on self-inductance compared to the impact on the quality factor. Consider, for instance, two spiral coils with a diameter of \SI{100}{\milli\meter}, a wire radius of \SI{1}{\milli\meter}, $5$ windings and a distance of \SI{100}{\milli\meter} between them. The coupling factor for \SI{100}{\kilo\hertz} and \SI{6.78}{\mega\hertz} estimated with the above equations taking the skin effect into account, results in $0.030561714$ and $0.030561708$ respectively. This confirms our expectations, and thus, the further results and insights of this section are applicable for the whole frequency band $\leq$~10~MHz.

The proposed model from \cref{subsubsection:math-k}, considering spiral coils, is used to investigate the impact of different dimensions and geometries of coils on the \gls{uav} on the coupling coefficient between the transmit and receive coils. For the transmit coil, the \gls{uav} leaves no room for adapting the coil diameter or number of windings; a fixed coil configuration is used at the transmit side. It should be noted that the shape of the transmit coil, on the \gls{uav}, more closely approximates a rectangular form compared to a spiral-shaped coil. To better match the dimensions of the \gls{uav}, the average of the x and y dimensions of the \gls{uav} model is considered as the radius of the outer concentric circle in \matlab{}. The distance between the concentric circles is assumed to be \SI{1}{\milli\meter}.
For the considered design, this results in radii $74.5$ and \SI{76.5}{\milli\meter}, knowing that there is space for two windings on the transmitter coil.
According to the \matlab{} model, this approximation of the transmit coil has a self-inductance of \SI{1.587}{\micro\henry}. For the receive coil at the \gls{iot} side there is more design freedom in order to optimize the coupling; various configurations are compared in the study.

In the next paragraphs we present the results of the analytical study of the coupling factor and in particular the impact of coil-to-coil distance (the vertical distance $\Delta Z$ between the coils), the different receive coil diameters and the lateral and angular misalignments. 

\begin{figure*}[h]
    \centering
\begin{tikzpicture}

\definecolor{darkgray176}{RGB}{176,176,176}
\definecolor{darkorange25512714}{RGB}{255,127,14}
\definecolor{steelblue31119180}{RGB}{31,119,180}

\begin{axis}[
width=0.8\textwidth,
height=0.3\textwidth,
axis background/.style={fill=gray!10},
tick align=outside,
tick pos=left,
x grid style={darkgray176},
xmajorgrids,
xmin=-8.95, xmax=209.95,
xtick style={color=black},
y grid style={darkgray176},
ylabel=\textcolor{steelblue31119180}{Coupling factor (k) [-]},
ymajorgrids,
ymin=0, ymax=1,
ytick style={color=black},
legend style={draw=none,yshift=-5cm,xshift=0.0cm, nodes={scale=1, transform shape}},
legend columns=2, 
]
\addlegendimage{semithick,draw=black,mark=*,mark options={black, solid}}
\addlegendentry{Receive coil diameter \SI{75}{\milli\meter}}
\addlegendimage{semithick,dashed,draw=black,mark=o,mark options={black, solid}}
\addlegendentry{Receive coil diameter \SI{100}{\milli\meter}}
\addlegendimage{semithick,dotted,draw=black,mark=triangle,mark options={black, solid}}
\addlegendentry{Receive coil diameter \SI{125}{\milli\meter}}
\addlegendimage{semithick,loosely dashdotted,draw=black,mark=diamond*,mark options={black, solid}}
\addlegendentry{Receive coil diameter \SI{150}{\milli\meter}}

\addplot [semithick, draw=steelblue31119180, mark options={color=steelblue31119180}, mark=*]
table{%
x  y
1 0.147390714
50 0.075932284
100 0.027916606
150 0.011778231
200 0.005789029
};
\addplot [semithick,dashed,draw=steelblue31119180, mark options={color=steelblue31119180, solid}, mark=o]
table{%
x  y
1 0.245296167
50 0.110924996
100 0.04028655
150 0.017253615
200 0.008581866
};
\addplot [semithick,dotted,draw=steelblue31119180, mark options={color=steelblue31119180, solid}, mark=triangle]
table{%
x  y
1 0.396725941
50 0.142531452
100 0.051778514
150 0.0226555
200 0.0114453
};
\addlegendimage{draw=black,mark=diamond*,mark options={black}}
\addplot [semithick,loosely dashdotted,draw=steelblue31119180, mark options={color=steelblue31119180, solid}, mark=diamond*]
table{%
x  y
1 0.775946757
50 0.165203676
100 0.061508011
150 0.027688345
200 0.014255924
};
\end{axis}

\begin{axis}[
width=0.8\textwidth,
height=0.3\textwidth,
axis y line=right,
tick align=outside,
x grid style={darkgray176},
xlabel={Coil-to-coil distance [\si{\milli\meter}]},
xmajorgrids,
xmin=-8.95, xmax=209.95,
xtick pos=left,
xtick style={color=black},
y grid style={darkgray176},
ylabel=\textcolor{darkorange25512714}{Self-inductance $L_2$ [$\mu H$]},
ymajorgrids,
ymin=0, ymax=6,
ytick pos=right,
ytick style={color=black},
yticklabel style={anchor=west}
]
\addplot [semithick,draw=darkorange25512714, mark options={color=darkorange25512714, solid}, mark=o]
table{%
x  y
1 1.8628
50 1.8932
100 1.9027
150 1.9039
200 1.9041
};
\addplot [semithick,dashed,draw=darkorange25512714, mark options={color=darkorange25512714, solid}, mark=o]
table{%
x  y
1 2.7314
50 2.8705
100 2.9016
150 2.9054
200 2.9061
};
\addplot [semithick,dotted,draw=darkorange25512714, mark options={color=darkorange25512714, solid}, mark=triangle]
table{%
x  y
1 3.3518
50 3.897
100 3.9672
150 3.9758
200 3.9773
};
\addplot [semithick,loosely dashdotted,draw=darkorange25512714, mark options={color=darkorange25512714, solid}, mark=diamond*]
table{%
x  y
1 2.0308
50 4.9643
100 5.0843
150 5.0997
200 5.1026
};
\end{axis}

\end{tikzpicture}
    \caption{Variation of the coupling coefficient and receiver coil self-inductance relative to the coil-to-coil distance. We note that variations in coil-to-coil distance induce also changes in the self-inductance of the transmitter coil, especially at short distances (i.e. for high coupling).}
    \label{fig:analysis-rx-coil}
\end{figure*}
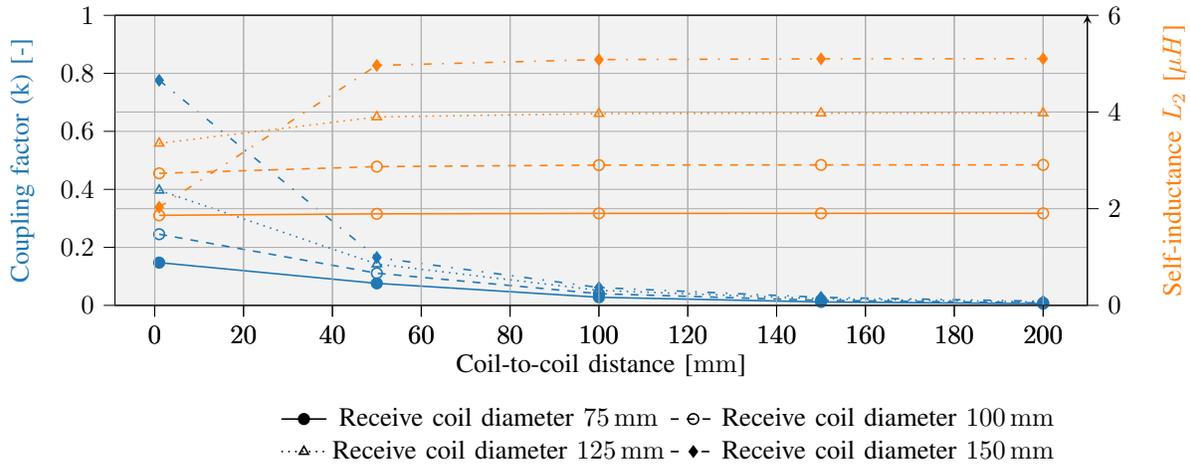

\paragraph{Impact of the coil-to-coil distance.} In situations with little or no wind, the coil-to-coil and lateral distances between transmit and receive coil will be the main factors affecting the coupling coefficient. A first assessment is performed, regarding coil-to-coil distance for four different dimensions of receive coils, each with four turns. 
The outer dimensions of the considered coils are \SI{75}{\milli\meter}, \SI{100}{\milli\meter}, \SI{125}{\milli\meter}, and \SI{150}{\milli\meter}. The impact of the coil-to-coil distance between transmit and receive coil, all assuming lateral alignment, is shown in \cref{fig:analysis-rx-coil}, and presents the impact on the coupling coefficients and the receiver coil self-inductance.

Upon reviewing \cref{fig:analysis-rx-coil}, it is observed that the coupling coefficient is the highest for the largest receive coil. This is as expected seeing that these dimensions better correspond to the dimensions of the transmit coil. Higher coupling factors, in general, allow for better efficiencies. However, the better the coupling, the higher the impact on the self-inductance. Due to the dynamic nature of a hovering \gls{uav}, variations in coil distances are likely over time. This could result in detuning of the coils and thus a drop in the power transfer efficiency, as further explained in \cref{subsection:mrcremarks}. 

For the remainder of this manuscript, a receiver coil with a \SI{100}{\milli\meter} diameter is preferred over the \SI{75}{\milli\meter} coil due to the slightly higher achievable coupling factors. \Cref{fig:analysis-rx-coil} demonstrate that the self-inductance of the \SI{100}{\milli\meter} receiver coil undergoes minimal changes and at the same time ensures adequate coupling upto \SI{100}{\milli\meter}.

Another trade-off to consider is the size of the \gls{iot} node and consequently, the size of the receive coil. Since the results indicate that sufficient coupling can be achieved with all proposed receiving coils, the smaller \SI{100}{\milli\meter} coil is selected to ensure practical feasibility.

\begin{figure}[!ht]
    \centering
    \begin{subfigure}[b]{0.45\textwidth}
        \centering
        \includegraphics[width=\textwidth]{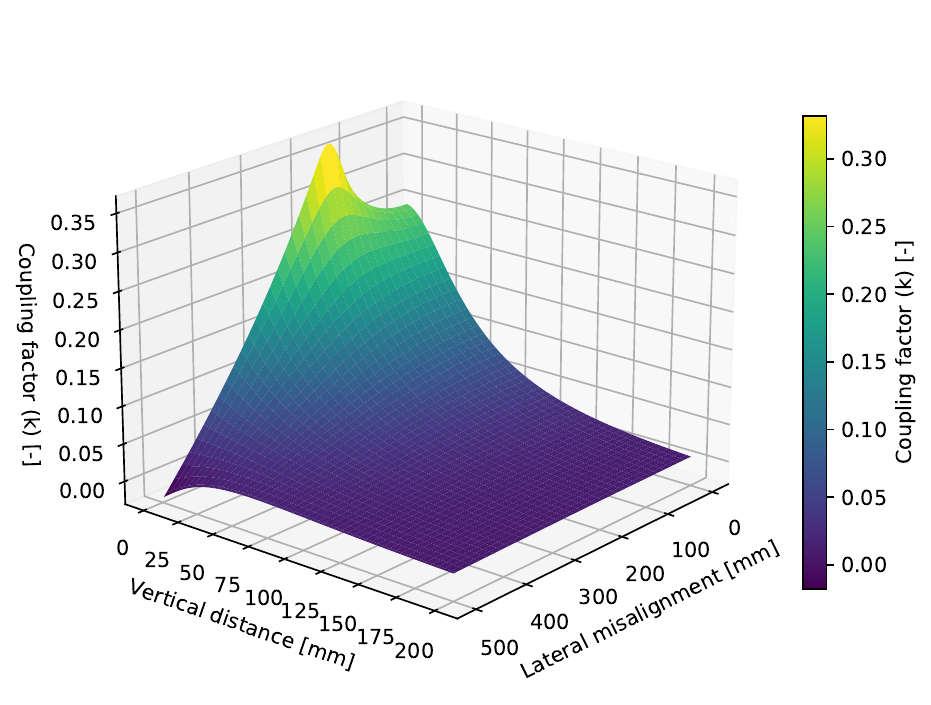}
        \caption{Increasing coil-to-coil distances and lateral misalignments.}
        \label{fig:vertical-lateral-misalignments}
    \end{subfigure}
    \begin{subfigure}[b]{0.45\textwidth}
        \centering
        \includegraphics[width=\textwidth]{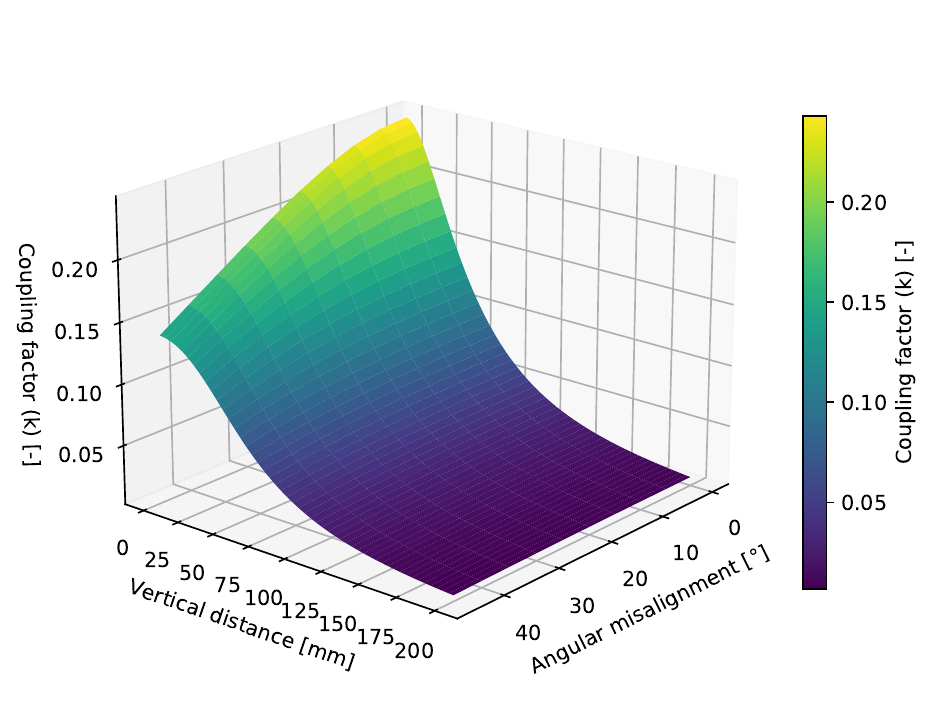}
        \caption{Increasing coil-to-coil distances and angular misalignments.}
        \label{fig:coupling-angular-misalignments}
    \end{subfigure}
    \caption{Coupling factor variations for different misalignments and with a receiver coil of \SI{100}{\milli\meter}.}
    \label{fig:subfigures}
\end{figure}

\paragraph{Coil-to-coil and lateral misalignments.} \label{para:c2c} 
\cref{fig:vertical-lateral-misalignments} a shows the impact of both coil-to-coil and lateral misalignments. This clearly indicates that the issues related to relative positioning can mutually amplify. For UAV-based charging, this implies that the transfer range between the node and drone is limited, and lateral misalignments must be kept within a few centimeters.

\paragraph{Number of coil windings.} \label{para:nowindings} The impact of the number of coil windings on the coupling factor was also assessed. The results, listed in \cref{tab:analysis-rx-coil-windings}, show that the impact on the coupling factor is rather low ($\leq$ 10\%). The number of windings has an impact on the self-inductance of the coil. The quantity can be chosen based on the desired self-inductance value to facilitate tuning to the working frequency, for instance, \SI{6.78}{\mega\hertz}.

\begin{table*}[h]
    \centering
    \resizebox{\textwidth}{!}{
    \begin{tabular}{c||cccccc||c||cc}
        \toprule
        Coil diameter & Windings & \multicolumn{5}{c||}{Radii $r$} & $L$ & \multicolumn{2}{c}{Coupling $k$} \\
        \,[\si{\milli\meter}] & & $r_{t_1}$ [\si{\milli\meter}] & $r_{t_2}$ [\si{\milli\meter}] & $r_{t_3}$ [\si{\milli\meter}] & $r_{t_4}$ [\si{\milli\meter}] & $r_{t_5}$ [\si{\milli\meter}] & [\si{\micro\henry}] & ($\Delta Z$ = \SI{50}{\milli\meter}) & ($\Delta Z$ = \SI{100}{\milli\meter}) \\       
        \midrule
            100 & 2 & 49.0 & 47.0 &      &      &      & 0.8998 & 0.1075 & 0.0390   \\
            100 & 3 & 49.0 & 47.0 & 45.0 &      &      & 1.806 & 0.1096 & 0.0398    \\ 
            100 & 4 & 49.0 & 47.0 & 45.0 & 43.0 &      & 2.906 & 0.1109 & 0.0403    \\
            100 & 5 & 49.0 & 47.0 & 45.0 & 43.0 & 41.0 & 4.145 & 0.1117 & 0.0406    \\
        \bottomrule
    \end{tabular}}
    \caption{Coupling between transmit and receive coil considering different number of windings at coil-to-coil distances $\Delta Z$ of 50 mm and 100 mm.}
    \label{tab:analysis-rx-coil-windings}
\end{table*}

\paragraph{Impact of angular misalignment.} When the two coils are not perfectly parallel the coupling factor will be influenced, this is depicted in \cref{fig:coupling-angular-misalignments}b. It is clear that for the \gls{uav}-based charging, a perfect angular positioning can not be guaranteed in case of wind for instance. This may cause a certain loss in actual power transfer. However, we may assume the angular misalignment error in general to be relatively small, typically $\leq$~10°. Hence, this will not significantly affect the coupling coefficient.

\subsection{MRC system introduction and mathematics}
\label{subsection:mrc-intro}

\cref{subsection:coupling-prediction} demonstrates that the coupling factors are expected around or lower than $0.1$, making \gls{mrc} the most suitable \gls{wpt} technology for these loosely coupled coils. This technology exhibits greater tolerance to changes in the coupling factor compared to other \gls{wpt} solutions while maintaining a sufficiently high power density over the link~\cite{van2022wireless}. The \gls{mrc} architecture was already shown in \cref{fig:high-level-overview}, whereby the \textit{\acrshort{wpt} transmitter} can be simplified as an \gls{ac} source with amplitude $V_S$ to drive the primary LC tank, and the \textit{\acrshort{wpt} receiver} gets its energy from the secondary LC tank and can be represented by an equivalent \gls{ac} load $R_L$. The \gls{ac} source is located at the \gls{uav}, while $R_L$ represents the charging circuit and energy buffer of the \gls{iot} node. A more comprehensive discussion concerning the design components can be found in \cref{section:design}. The section focuses on the \gls{mrc} system and related implementation choices.

In \gls{mrc} both transmitter and receiver are perfectly tuned to the resonance frequency as given in \cref{eq:resonance}. The \gls{ism} frequency \SI{6.78}{\mega\hertz} is selected as carrier. This relatively high frequency, compared to the often applied \SI{100}{\kilo\hertz} (in Qi~\cite{wpcqispecpd}), ensures the use of lower cost \gls{pcb} coils, lower heating problems or eddy currents, and higher quality factors between transmit and receive coils. Given the thin \gls{pcb} traces, typically \SI{35}{\micro\meter} thick, the skin effect can be disregarded and no expensive Litz wire is required.

\begin{equation}
    \omega_0 = \cfrac{1}{\sqrt{L_1 \cdot C_1}} = \cfrac{1}{\sqrt{L_2 \cdot C_2}}
    \label{eq:resonance}
\end{equation}

Series-series resonance is the preferred tuning configuration, particularly due to its higher efficiency compared with other options. In \cite{imura2020wireless}\,(Ch.\,5) provides a comprehensive elaboration on the fundamental tuning types: N-N, N-S, S-N, S-S, and S-P, where N, S, and P represent non-resonant, series resonant, and parallel resonant transceiver tank, respectively. This study reveals that the highest efficiencies can be achieved with S-S and S-P configurations. S-S is equivalent to a current source at the secondary, while S-P is equivalent to a voltage source at the secondary. In both circuit types, $C_1$ is responsible for facilitating high power transfer, and $C_2$ is instrumental in achieving the required efficiency levels. For the remainder of this paper, we assume the S-S configuration as depicted in \cref{fig:high-level-overview}.

Our objective is to determine the link efficiency related to the power consumed in the load $R_L$. Obviously, having an infinitely small or an infinitely large load, will results in zero efficiency. Hence, there exists an optimal point $R_{L,opt}$. Furthermore, it can be advantageous to estimate the input voltage $V_S$ based on the required load power $P_{L}$. To derive the expression for the link efficiency and determine the optimal parameters, the circuit diagram depicted in \cref{fig:high-level-overview} can be substituted with the equivalent T-model network, and Kirchhoff's voltage law can be applied. The non-ideal coils $L_1$ and $L_2$ incorporate series resistances $R_{1}$ and $R_{2}$, respectively. The internal series resistor from the power supply $V_S$ is represented by $R_S$. 

The link efficiency determined by the power consumed in the load $P_{L}$ relative to the input power $P_S$ is given by \cref{eq:mrc-eff}.

\begin{equation}
    \eta_{link} = \frac{P_L}{P_S} = \frac{R_L}{R_{L_2} + R_L}\frac{k_{mrc}^2 Q_T Q_R}{1 + k_{mrc}^2 Q_T Q_R}
    \label{eq:mrc-eff}
\end{equation}

with $k_{mrc}$ the coupling factor and the quality factor of transmitter and receiver circuit, $Q_T$ and $Q_R$, as specified in \cref{eq:q-factors}.

\begin{equation}
    \begin{aligned}
        Q_T = \cfrac{\omega_0 L_1}{R_S + R_{1}}\,\,,\,\,Q_R = \cfrac{\omega_0 L_2}{R_{2} + R_L}\\
        Q_1 = \cfrac{\omega_0 L_1}{R_{1}}\,\,,\,\,Q_2 = \cfrac{\omega_0 L_2}{R_{2}}
    \end{aligned}
    \label{eq:q-factors}
\end{equation}

The maximum \gls{mrc} efficiency will occur for a specific value of $R_L$ and can be determined by taking the derivative of formula \cref{eq:mrc-eff}, as given in \cref{eq:mrc-optimal-load}.

\begin{equation}
    \begin{aligned}
        \frac{d(\eta_{link})}{d(R_L)} = 0 \\
        \rightarrow\,\, R_{L,opt} = \sqrt{R_2^2 \cdot \left(1 + k_{mrc}^2 Q_1 Q_2 \frac{R_1}{R_S + R_1} \right)}
    \end{aligned}
    \label{eq:mrc-optimal-load}
\end{equation}

For a more detailed derivation, we direct the reader to~\cite{van2022wireless,ahn2012study,baker2007feedback,rindorf2008resonantly}.

\subsection{Concluding remarks regarding the \Gls{mrc} system design for \gls{uav}-based charging}
\label{subsection:mrcremarks}

The default procedure in the design of the \gls{mrc} charging system would be to select transmitter and receiver coils and measure the self-inductance in open air. Consequently, the LC tanks are tuned to a certain frequency band by adding an appropriate series capacitor. Maximum transfer is guaranteed if LC tuning is optimal. However, \cref{fig:analysis-rx-coil} shows that $L_1$ and $L_2$ relate to the vertical distance between the coils. Thus, positioning the coils in close proximity to each other can alter the inductance $L_1$ and $L_2$, potentially causing a mismatch between the tuning frequency and the designed operating frequency, thereby no longer satisfying \cref{eq:resonance}. Hence, it is imperative to avoid excessive coupling as its occurrence will inevitably lead to a decline in efficiency. As \cref{fig:analysis-rx-coil} shows, appropriate coil selection can avoid this.

\section{WPT design for UAV-based charging}
\label{section:coil-design}

Before proceeding with the implementation, we considered it important to assess the expected  efficiency levels with a higher confidence level through more detailed simulations and calculations. This chapter is dedicated to a thorough exploration of the essential building blocks required for the \gls{wpt} system, including the selection of the coils, and the determination of the maximum achievable link efficiency.

\subsection{Coil design}
\label{subsection:coil-design}

To pursue a more in-depth analysis of the proposed system, the coils need to be defined initially. The transmitter coil shape must align with the \gls{uav} model presented in \cref{fig:uav-iot-full-render} and has dimension $146$\,×\,\SI{164}{\milli\meter}. From the initial analysis in \cref{subsection:coupling-prediction}, it has been determined that a receiver coil with four windings and dimensions of $100$\,×\,\SI{100}{\milli\meter} is a suitable choice. It could pick up the alternating magnetic field energy when the \gls{uav} is approaching a depleted \gls{iot} device as specified in \cref{fig:analysis-rx-coil}. Authors in~\cite{hackl2008novel} describe how to measure the coupling coefficient $k_{mrc}$ via different approaches and comment on the parasitic capacitive coupling between closely coupled coils. Since the distance between transmitter and receiver coils is sufficiently high in the application at hand, the capacitive coupling can be ignored~\cite{jeon2019coupling}. The electrical coupling is negligible compared to the magnetic coupling.

A \gls{fem} analysis is presented in \cref{subsubsection:fem}, providing the coupling factor variation as a result of vertical and lateral misalignment. These simulations are conducted using only coils in free space, without considering any neighboring materials. In \cref{subsubsection:two-port}, the results from the \gls{fem}-based assessment are validated with actual measurements, and the impact of additional \gls{uav} elements in the neighborhood of the transmit coil on the coupling factor is evaluated. This is accomplished using the two-port method detailed in \cite{jeon2019coupling}. \cref{subsubsection:tuning} validates the tuning quality through a one-port measurement.

\subsubsection{\Gls{fem} simulation to define variations in coupling factor}
\label{subsubsection:fem}

In the Ansys Maxwell \acrlong{fem} software tool~\cite{ansys-maxwell}, the two coil models were created to investigate how coil parameters and, consequently, the coupling factor are influenced by different misalignments. The coils are modeled using an FR4 PCB substrate (2-layer coil) with a thickness of \SI{0.6}{\milli\meter}. The \gls{fem} tool simulates the self-inductance $L_1$, equivalent series resistance $R_1$, the self-inductance $L_2$, equivalent series resistance $R_2$ and mutual inductance $M_{12}$. Knowing that the frequency is \SI{6.78}{\mega\hertz}, the coupling factor $k$ and the quality factors $Q_1$ and $Q_2$ can be determined using \cref{eqn:k,eq:q-factors}. Several results from the \gls{fem} analysis regarding the coupling factor are depicted in \cref{tab:simmeascouplingfactor}.

\subsubsection{Measurement-based validation of the computed coupling factors}
\label{subsubsection:two-port}

The finite element simulations from \cref{subsubsection:fem} are verified by measuring the coupling factor between the coils with the characteristics as used in the simulations. Since modelling a \gls{uav} is rather complex, it is here preferred to measure the influence of the \gls{uav} components and materials on the coupling factor. As explained in \cite{jeon2019coupling}, the coupling factor can be determined via a two port measurement with a \gls{vna}. Before performing the measurements, the R\&S VNA ZVL3 and a coaxial cable are calibrated with the Siglent F604FS 9 GHz SMA-Female calibration kit. The experimental configuration is similar to \cref{fig:meas-setup}, differing only in that solely the coils were interfaced with the \gls{vna}. Notably, the additional electronic components depicted in this figure are superfluous for the purpose of determining the coupling factor. Via \gls{scpi}, the \gls{vna} returns the [S] matrix which can be converted to the [Z] matrix (represented in \cref{eq:z-matrix}) and ultimately the coupling factor.

\begin{equation}
    \begin{bmatrix}Z\end{bmatrix} = 
    \begin{bmatrix}
      Z_{11} & Z_{12} \\
      Z_{21} & Z_{22} \\
    \end{bmatrix}
    \label{eq:z-matrix}
\end{equation}

The equation to convert the measurement data to $k_{mrc}$ is presented in \cref{eqn:k} with $L_1$, $L_2$, and $M_{12}$ corresponding to $\operatorname{Im}(Z_{11})$, $\operatorname{Im}(Z_{22})$, and $\operatorname{Im}(Z_{12})$, respectively~\cite{jeon2019coupling}. The results for multiple vertical misalignments are represented in \cref{tab:simmeascouplingfactor}. The measurement setup and corresponding scripts can be accessed from our repository~\cite{vnaCouplingMeas}.

\begin{table}[h]
    \centering
    \resizebox{0.5\textwidth}{!}{
    \begin{tabular}{ccc||c||c||cc}
        \toprule
        \multicolumn{3}{c||}{Misalignment} & \matlab{} & \Gls{fem} & \multicolumn{2}{c}{Two port measurement} \\
        $\Delta X$ & $\Delta Y$ & $\Delta Z$ &  &  & Open air & On UAV \\
        \,[\si{\milli\meter}] & [\si{\milli\meter}] & [\si{\milli\meter}] & Coupling [-] & Coupling [-] & \multicolumn{2}{c}{Coupling [-]} \\
        \midrule
          0 & 0 & 50   & 0.111 & 0.094 & 0.107 & 0.113 \\
          0 & 0 & 100  & 0.040 & 0.035 & 0.042 & 0.044 \\
          0 & 0 & 150  & 0.017 & 0.014 & 0.018 & 0.018 \\
          0 & 0 & 200  & 0.009 & 0.005 & 0.010 & 0.011 \\
        \bottomrule
    \end{tabular}}
    \caption{Coupling factor $k$ simulated and measured for different vertical misalignment distances.}
    \label{tab:simmeascouplingfactor}
\end{table}

\Cref{tab:simmeascouplingfactor} reveals that the spiral coils in the \matlab{} calculations serve as a reliable approximation for the practical PCB coils. The results from \matlab{}, \gls{fem}, and two-port network measurements all demonstrate consistency. We can see that mathematical calculations are a suitable method for initial coupling coefficient predictions. A final measurement using a \gls{vna} remains the most reliable method to verify the influences of the surroundings. In accordance with the findings from this final method, it is evident that the surrounding \gls{bldc} motors and controller board exhibit a negligible impact on the coupling coefficient in our case.

\subsubsection{Coil tuning capacitors}
\label{subsubsection:tuning}

\begin{figure*}[h]
    \centering
    \includegraphics{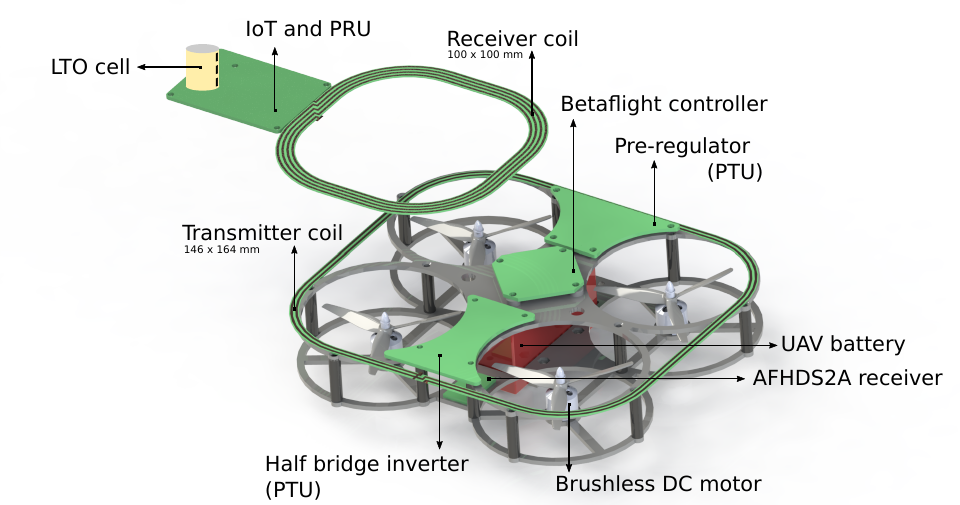}
    \caption{Render of the full system for the proposed approach to charge \acrshort{iot} nodes autonomously. The \acrshort{uav} battery delivers \acrshort{dc} power to the Betaflight controller and pre-regulator. The latter forwards power to the half-bridge inverter and is further connected to the transmission coil. The receiver coil receives the alternating magnetic field and charges the \acrshort{iot} battery.}
    \label{fig:uav-iot-full-render}
\end{figure*}

As discussed in \cref{subsection:mrc-intro}, \gls{mrc} systems are constructed with two perfectly tuned coils. This application uses the working frequency of \SI{6.78}{\mega\hertz}. With a calibrated 1~port \gls{vna}, the initial transmitter impedance of $0.1+j84$ \si{\ohm} was adapted to an impedance of $0.6+j2.7$ \si{\ohm} by a series resonance capacitor of \SI{276}{\pico\farad} (\SI{220}{\pico\farad} + \SI{56}{\pico\farad}). Similarly, the receiver coil with an impedance of $1+j143$ \si{\ohm} was matched with a series capacitor of \SI{135}{\pico\farad}, giving an impedance of approximately $1+j5$ \si{\ohm}. In both cases the reactive component is nearly completely neutralized by the series capacitor. In the final setup, the transmitter coil is surrounded by various printed circuit boards as depicted in \cref{fig:uav-iot-full-render}. The measured impedance is approximately $1+j3$ \si{\ohm} when the coil is mounted on the UAV, indicating that the impact of the mounting is relatively small, and, in fact, it even improves the tuning in this specific implementation.

\subsection{Achievable wireless power transfer efficiency}
\label{subsection:wpt-link-eff}

\begin{figure}[!ht]
    \centering
    \begin{subfigure}[b]{0.45\textwidth}
        \centering
        \includegraphics[width=\textwidth]{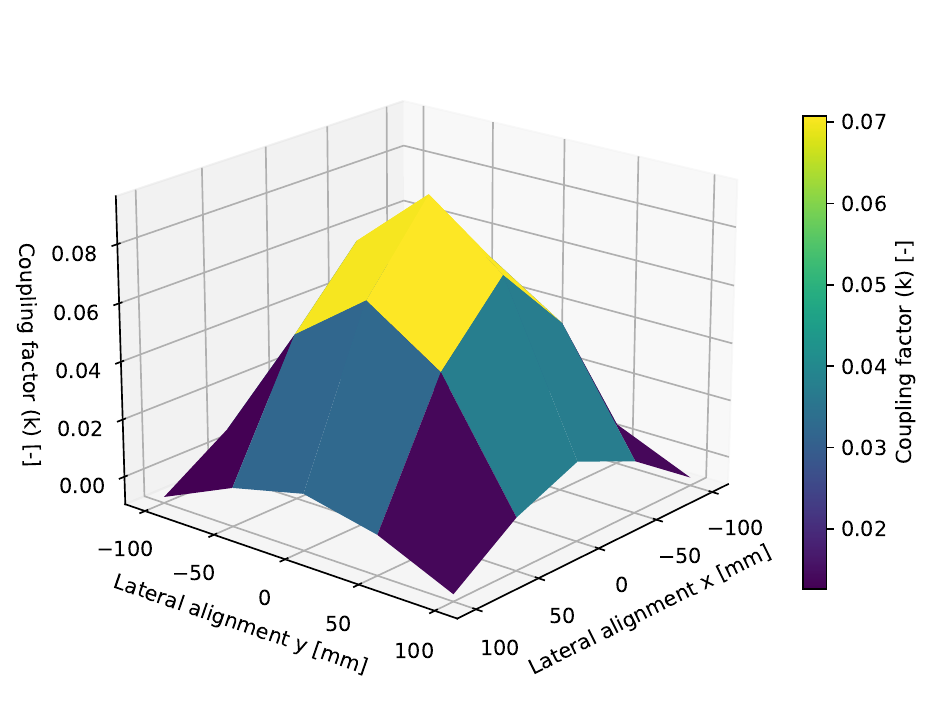}
        \caption{Coil-to-coil distance \SI{50}{\milli\meter}}
        \label{fig:max-link-50}
    \end{subfigure}
    \begin{subfigure}[b]{0.45\textwidth}
        \centering
        \includegraphics[width=\textwidth]{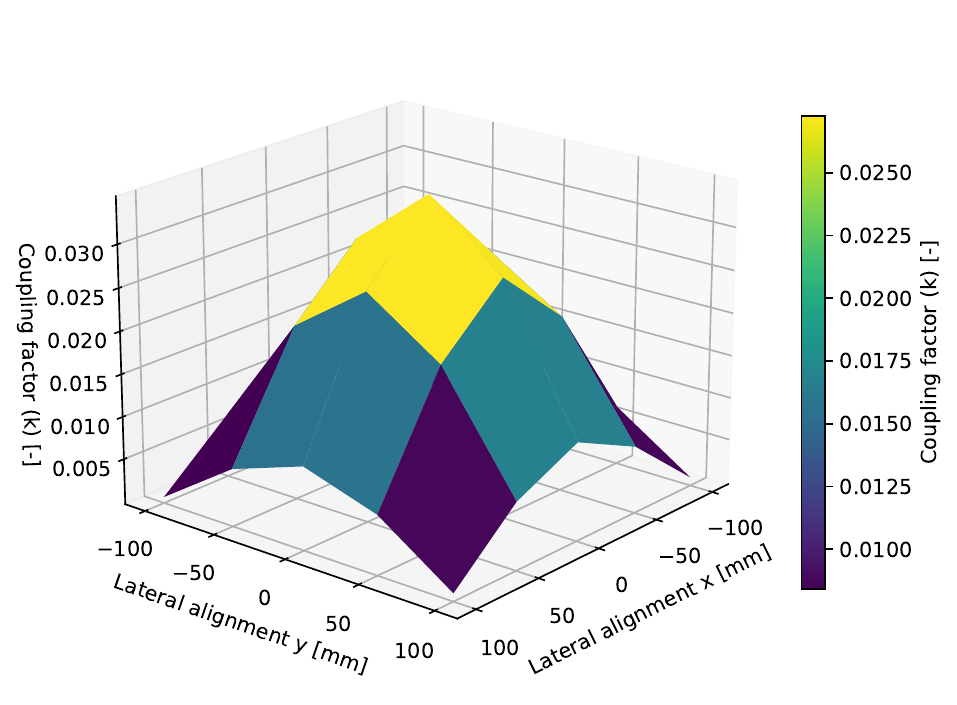}
        \caption{Coil-to-coil distance \SI{100}{\milli\meter}}
        \label{fig:max-link-100}
    \end{subfigure}
    \caption{Graphical representation of the maximum attainable link efficiency, as derived from \gls{fem} results, under optimal load conditions.}
    \label{fig:couplingfactor}
\end{figure}

The maximum achievable link efficiency can be calculated using \cref{eq:mrc-eff}. To achieve this optimal coil-to-coil efficiency, the link is assumed to operate at the optimal load, which can be calculated based on \cref{eq:mrc-optimal-load}. The \gls{fem} results are utilized to determine $\eta_{link_{max}}$, where each data point with a misalignment ($\Delta X$, $\Delta Y$, $\Delta Z$) includes $L_1$, $L_2$, $R_1$, $R_2$, $k_{mrc}$. Given these earlier \gls{fem} results, $R_{L,opt}$ can be computed and substituted into \cref{eq:mrc-eff}. \Cref{fig:couplingfactor} depict the maximum achievable \gls{wpt} link efficiencies for a coil-to-coil distance of $50$ and \SI{100}{\milli\meter}, respectively.

These results indicate that efficiencies across the WPT link can exceed \SI{70}{\percent} at coil-to-coil distances up to \SI{100}{\milli\meter}. However, as lateral distances increase, achievable efficiencies rapidly decline.

\section{Charging system design and implementation}
\label{section:design}

The design of the \gls{uav}-based charging system can pay attention to achieve the maximum attainable link efficiency as discussed in \cref{subsubsection:tuning}. Furthermore, other blocks are also necessary to transfer energy from the \gls{uav} battery to the \gls{lto} cell. These additional blocks are discussed in \cref{subsubsection:wpt-building-blocks}, further zooming in on the receiver and transmitter units (in \cref{subsection:pru-design} and \cref{subsection:ptu-design} respectively), and the data communication (\cref{subsection:Data_com}). 

\subsection{WPT building blocks}
\label{subsubsection:wpt-building-blocks}

The transmitter depicted in \cref{fig:high-level-overview} includes separate blocks to actually enable \acrlong{wpt} and act as the \gls{ac} source. The \gls{uav} battery is connected to a pre-regulator, essentially a DC-DC converter, allowing the voltage to be dynamically controlled by the processing unit of the \gls{wpt} transmitter. This variable voltage is then connected to the power inverter. The inverter transforms the pre-regulator DC voltage into a square wave signal with a frequency of \SI{6.78}{\mega\hertz}. Subsequently, a \gls{zvs} circuit is placed between the inverter output and the primary LC tank. These blocks are further explained in \cref{subsection:ptu-design}.

On the receiver side, the \gls{iot} node, the alternating magnetic field is captured by the receiver LC tank and then rectified to a DC voltage. This AC-DC conversion comes with associated losses, where the diode threshold voltage determines the efficiency. Typically, Schottky diodes are selected due to their lower forward voltage levels. The last crucial element is the DC-DC converter, which ensures the charging process of the \gls{lto} cell. The DC-DC converter is set up as a \gls{cc}/\gls{cv} power supply. The $R_L$ in \cref{fig:high-level-overview} thus encompasses the AC-DC conversion, the DC-DC \gls{cc}/\gls{cv} charger, and the \gls{lto} cell itself. These additional electronic components are required to make an \gls{iot} device compatible with these approach and thus convert the alternating magnetic fields to energy and consequently charge the battery.

\Cref{fig:uav-iot-full-render} presents a full render of all components to enable wireless power transfer and ensure the \gls{uav} can fly.


\subsection{\Acrfull{pru}}
\label{subsection:pru-design}
The \gls{iot} device and \gls{pru} presented in \cref{fig:uav-iot-full-render} have been integrated into a single unit and are connected with (1) a receiver coil (2) a \gls{lpwan} modem (e.g. \gls{lora} connection) and (3) one or more sensors or actuators. Therefore, having an onboard \gls{mcu} to communicate with the sensors and \gls{lpwan} modem is needed. A nordic NRF52832 \gls{sysoc} with \gls{ble} functionality has been selected and is utilized to control both the charging process and \gls{iot}-related tasks. During the charging process,  a \gls{ble} link between the \gls{iot} device and the \gls{uav} is established to facilitate real-time communication of input voltages, charge voltages and charge currents. The \gls{ptu} can respond to this measurement data by, for example, fine-tuning the alignment or adjusting the pre-regulator voltage. During normal \gls{iot} operation, the \gls{sysoc} can manage \gls{iot}-related tasks, such as capturing and processing sensor data and forwarding the results via the \gls{lpwan} modem to the cloud.

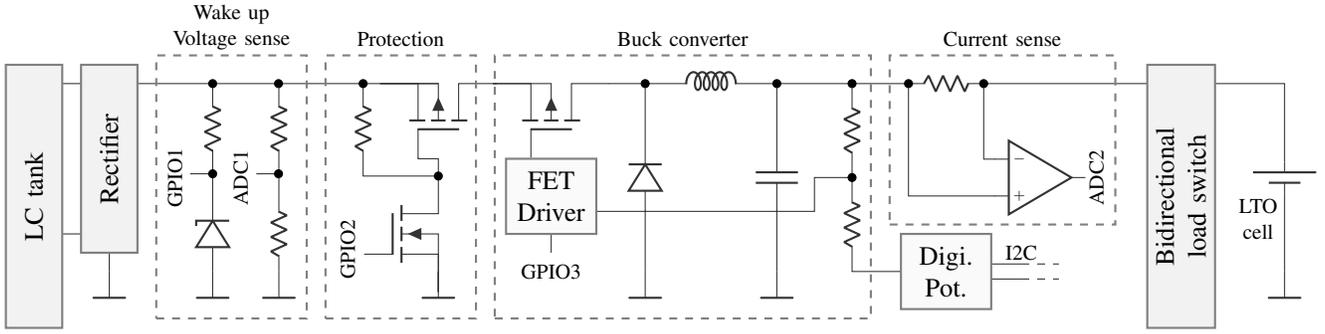
\begin{figure*}
    \centering


\begin{tikzpicture}[american voltages, every text node part/.style={align=center}, /tikz/circuitikz/tripoles/nigfetd/height=.8, /tikz/circuitikz/tripoles/nigfetd/width=.5]

\ctikzset{bipoles/resistor/height=0.15}
\ctikzset{bipoles/resistor/width=0.4}
\ctikzset{bipoles/capacitor/height=0.4}
\ctikzset{bipoles/capacitor/width=0.1}
\ctikzset{bipoles/americaninductor/height=0.35}
\ctikzset{bipoles/americaninductor/width=0.8}

\ctikzset{resistor=american}

\tikzset{block/.style = {rectangle, draw=black!50, fill=black!5, thick, minimum width=3.5cm, minimum height = 0.75cm}}
\tikzset{rect/.style = {rectangle, draw=black!50, fill=black!5, thick, minimum width=2.5cm, minimum height = 0.75cm}}
\tikzset{block2/.style = {rectangle, draw=black!50, fill=black!5, thick, minimum width=2.2cm, minimum height = 3cm}}
\tikzset{blocksmall/.style = {rectangle, draw=black!50, fill=black!2, thick, minimum width=1.2cm, minimum height = 1cm}}
\tikzset{blocklong/.style = {rectangle, draw=black!50, thick, minimum width=2.5cm, minimum height = 0.75cm}}

\tikzset{init/.style = {rectangle, dashed, draw=black!50, thick, minimum width=2cm, minimum height=3.5cm}}
\tikzset{prot/.style = {rectangle, dashed, draw=black!50, thick, minimum width=2cm, minimum height=3.5cm}}
\tikzset{curr/.style = {rectangle, dashed, draw=black!50, thick, minimum width=3cm, minimum height=2.25cm}}
\tikzset{buck/.style = {rectangle, dashed, draw=black!50, thick, minimum width=5cm, minimum height=3.5cm}}

\tikzset{fontscale/.style = {font=\relsize{#1}}}

\draw[color=black!80]

(-2.375,1) -- (-2.375,0) node[rground]{}
(-3.5,0.5) -- (-2.5,0.5) 

(-1,0) node[rground]{}
(-1,0) to [zD, /tikz/circuitikz/bipoles/length=25pt](-1,1) -- (-1,1.3)
(-1,1.3) to [R, /tikz/circuitikz/bipoles/length=40pt](-1,2.5) -- (0,2.5)
(-3.5,2.5) -- (-1, 2.5)
(-1,1.3) -- ++(-0.3,0)

(-0.12,0) node[rground]{}
(-0.12,0) to [R, /tikz/circuitikz/bipoles/length=40pt](-0.12,1) -- (-0.12,1.3)
(-0.12,1.3) to [R, /tikz/circuitikz/bipoles/length=40pt](-0.12,2.5)
(-0.12,1.3) -- ++(-0.3,0)

(1,1.3) to [R, /tikz/circuitikz/bipoles/length=40pt](1,2.5)
(0,2.5) -- (2,2.5) node[pigfete,scale=1, rotate=90] (pmos1){}
(2,0.5) node[nigfete,scale=1] (nmos1) {} 
(1,1.3) |- (nmos1.D)
(pmos1.G) |- (2,1.5) -- (nmos1.D)
(nmos1.S) to (2,0) node[rground]{}

(3.5,2.5) node[pigfete,scale=1, rotate=90] (pmos2){}
(3.5,0.25) -- (3.5,0.5)
(4.75,0) node[rground]{}
(4.75,0) to [D, /tikz/circuitikz/bipoles/length=25pt](4.75,2.5)
(pmos2.D) -- (4.75,2.5) to [L, /tikz/circuitikz/bipoles/length=30pt] (6.5,2.5) -- (7.5,2.5)
(6.5,2.5) to [C] (6.5,0)
(6.5,0) node[rground]{}
(8.25,0) -- (7.5,0) to [R, /tikz/circuitikz/bipoles/length=40pt](7.5,1.25) -| (7,0.8) -- (4,0.8)
(7.5,1.25) to [R, /tikz/circuitikz/bipoles/length=40pt](7.5,2.5)
(9.25,0.1) -- (9.75,0.1)
(9.25,-0.1) -- (9.75,-0.1)

(7.5,2.5)--(8,2.5) to [R, /tikz/circuitikz/bipoles/length=40pt](9.5,2.5)
(10, 1.25) node[op amp, scale=0.5] (opamp) {}
(opamp.+) -| (8.25, 2.5)
(opamp.-) -| (9.25, 2.5)

(9.5,2.5) -- (13.25,2.5) to [battery1] (13.25,0)
(13.25,0) node[rground]{}

;

\draw[color=black!80, dashed]
(9.75,0.1) -- (10.25,0.1)
(9.75,-0.1) -- (10.25,-0.1)
;

\filldraw[black] (-1,1.3) circle (1.5pt) node[anchor=west]{};
\filldraw[black] (-1,2.5) circle (1.5pt) node[anchor=west]{};
\filldraw[black] (-0.125,1.3) circle (1.5pt) node[anchor=west]{};
\filldraw[black] (-0.125,2.5) circle (1.5pt) node[anchor=west]{};
\filldraw[black] (1,2.5) circle (1.5pt) node[anchor=west]{};
\filldraw[black] (nmos1.D) circle (1.5pt) node[anchor=west]{};
\filldraw[black] (4.75,2.5) circle (1.5pt) node[anchor=west]{};
\filldraw[black] (6.5,2.5) circle (1.5pt) node[anchor=west]{};
\filldraw[black] (7.5,1.25) circle (1.5pt) node[anchor=west]{};
\filldraw[black] (7.5,2.5) circle (1.5pt) node[anchor=west]{};
\filldraw[black] (8.25, 2.5) circle (1.5pt) node[anchor=west]{};
\filldraw[black] (9.25, 2.5) circle (1.5pt) node[anchor=west]{};

\node [scale=0.8, rotate=0] at (12.9,0.7) () {LTO\\cell};

\node [block, rotate=90] at (-3.375,1) () {LC tank};
\node [rect, rotate=90] at (-2.375,1.5) () {Rectifier};
\node [blocksmall, rotate=0] at (3.5,1) () {FET\\Driver};
\node [blocksmall, rotate=0] at (8.75,0) () {Digi.\\Pot.};

\node [init] at (-0.75,1.125) () {};
\node [prot] at (1.5,1.125) () {};
\node [buck] at (5.25,1.125) () {};
\node [curr] at (9.5,1.75) () {};

\node [scale=0.8, rotate=0] at (9.75,0.25) () {I2C};

\node [scale=0.8, rotate=90] at (-1.5,1.25) () {GPIO1};
\node [scale=0.8, rotate=90] at (-0.625,1.25) () {ADC1};
\node [scale=0.8, rotate=90] at (0.8,0.3) () {GPIO2};
\node [scale=0.8, rotate=0] at (3.5,0) () {GPIO3};

\node [scale=0.8, rotate=90] at (10.75,1.25) () {ADC2};

\node [scale=0.8, rotate=0] at (-0.75,3.25) () {Wake up\\ Voltage sense};
\node [scale=0.8, rotate=0] at (1.5,3.1) () {Protection};
\node [scale=0.8, rotate=0] at (5.25,3.1) () {Buck converter};
\node [scale=0.8, rotate=0] at (9.5,3.1) () {Current sense};

\node [block, rotate=90] at (11.875,1) () {Bidirectional\\ load switch};

\end{tikzpicture}%

    \caption{Schematic representation of charging circuit. GPIO1, GPIO2, and GPIO3 correspond to the wake-up input, protection enable output, and buck enable output, respectively. ADC1, and ADC2 enable the measurement of input voltage and output current. (\acrshort{mcu} and \acrshort{iot}-related parts not depicted here.)}
    \label{fig:receiver-charger}
\end{figure*}

This section further comprehensively describes the \gls{pru} depicted in \cref{fig:uav-iot-full-render}. This charger circuit features several properties required for a safe and reliable charge. It can be divided into four main features: the charging circuitry itself, input voltage protection, leakage current reduction and charge suppression. \Cref{fig:receiver-charger} represents the charging circuit.

The core of the charging circuit is a buck converter whose feedback loop voltage divider can be adjusted with a programmable potentiometer. The current can be reduced by adjusting the feedback loop and is measured with a shunt resistor and opamp. This analog voltage is fed to the \gls{mcu}. The latter can adjust the digital potentiometer via the I2C interface.

The \gls{fet} and \gls{fet} driver from the buck converter circuit in \cref{fig:receiver-charger} are integrated in a buck converter \gls{ic} and can handle input voltages up to \SI{28}{\volt}. In wireless power transfer systems, the open circuit voltage of the LC tank can exceed this maximum voltage barrier and could damage the charging circuitry. A load switch design with a sufficient high input range will only pass voltages if a save level is measured. An \gls{mcu}, powered by the \gls{lto} cell, measures the input voltage and drives the load switch to connect the LC tank voltage to the buck converter input to start the charging process.

Many battery-powered devices are charged with a wall adapter. In these devices, the quiescent currents generated by the connected adapter circuit do pose significant autonomy problems. The impact of quiescent currents, caused by the non-active charging circuit, for large battery capacities is much lower compared to small battery capacities. In the considered devices, the \gls{lto} cell has a low capacity. Therefore, the quiescent current should also be very low. An additional bidirectional load switch is provided that disconnects the charger from the \gls{lto} cell and permits the flow of current from the charger to the cell. This approach is preferred over using a diode due to the additional diode losses, voltage drops, and reverse diode currents. The predicted autonomy, in the absence of any application, or \gls{lto} self-discharge is determined by dividing the available capacity (\SI{60}{\milli\ampere\hour}) by the leakage current (\SI{1.5}{\micro\ampere}), resulting in an estimated operational duration of approximately \SI{4.5}{\years}.

If the rectifier voltage is too low, it indicates a suboptimal coil alignment. A buck converter supplied with insufficient input voltage cannot operate effectively, necessitating reduced charging. This charge suppression becomes active at a rectified voltage lower than \SI{10}{\volt} and is achieved by adjusting the digital potentiometer. This adjusted resistance in the feedback loop of the buck converter leads to a reduction in output voltage and, consequently, output current.

\begin{figure*}[h]
    \centering
\begin{tikzpicture}

\definecolor{crimson2143940}{RGB}{214,39,40}
\definecolor{darkgray176}{RGB}{176,176,176}
\definecolor{darkorange25512714}{RGB}{255,127,14}
\definecolor{forestgreen4416044}{RGB}{44,160,44}
\definecolor{lightgray204}{RGB}{204,204,204}
\definecolor{steelblue31119180}{RGB}{31,119,180}

\begin{axis}[
width=0.8\textwidth,
height=0.4\textwidth,
axis background/.style={fill=gray!10},
legend cell align={left},
legend style={
  fill opacity=0.8,
  draw opacity=1,
  text opacity=1,
  at={(0.03,0.03)},
  anchor=south west,
  draw=lightgray204
},
tick align=outside,
tick pos=left,
x grid style={darkgray176},
xlabel={Input voltage sweep [V]},
xmajorgrids,
xmin=7.35, xmax=21.65,
xtick style={color=black},
y grid style={darkgray176},
ylabel={Efficiency [\%]},
ymajorgrids,
ymin=48, ymax=82,
ytick style={color=black}
]
\addplot [draw=steelblue31119180, fill=steelblue31119180, forget plot, mark=*, only marks]
table{%
x  y
8 76.4010989010989
9 74.4578313253012
10 73.1842105263158
11 71.2163892445583
12 69.1791044776119
13 66.8509615384615
14 65.1288056206089
15 62.8474576271186
16 59.9353448275862
17 58.4243697478992
18 55.1785714285714
19 53.2248803827751
20 51.5
21 49.9730458221024
};
\addplot [draw=darkorange25512714, fill=darkorange25512714, forget plot, mark=*, only marks]
table{%
x  y
8 78.0113636363636
9 78.7383512544803
10 77.9007092198582
11 76.8111888111888
12 76.2777777777778
13 74.7719537100068
14 74.0161725067386
15 71.7908496732026
16 70.7731958762887
17 68.7359198998748
18 67.8024691358025
19 65.6937799043062
20 64.6117647058824
21 62.2675736961451
};
\addplot [draw=forestgreen4416044, fill=forestgreen4416044, forget plot, mark=*, only marks]
table{%
x  y
8 76.9602272727273
9 77.5107296137339
10 78.1442307692308
11 77.7703349282297
12 77.4
13 77.1794871794872
14 76.3815789473684
15 75.25
16 74.1514598540146
17 73.5475113122172
18 71.6666666666667
19 70.7003044802088
20 69.4615384615385
21 67.8947368421052
};
\addplot [draw=crimson2143940, fill=crimson2143940, forget plot, mark=*, only marks]
table{%
x  y
8 75.0561797752809
9 76.6164874551971
10 77.1696750902527
11 77.4212241941326
12 77.4492753623188
13 76.8368080517613
14 76.7264895908112
15 75.8014184397163
16 75.0561797752809
17 74.4030630003481
18 73.3058984910837
19 72.5840407470289
20 71.7315436241611
21 70.6878306878307
};
\addplot [semithick, steelblue31119180]
table {%
8 76.4010989010989
9 74.4578313253012
10 73.1842105263158
11 71.2163892445583
12 69.1791044776119
13 66.8509615384615
14 65.1288056206089
15 62.8474576271186
16 59.9353448275862
17 58.4243697478992
18 55.1785714285714
19 53.2248803827751
20 51.5
21 49.9730458221024
};
\addlegendentry{\tiny0.2 A (0.56 W)}
\addplot [semithick, darkorange25512714]
table {%
8 78.0113636363636
9 78.7383512544803
10 77.9007092198582
11 76.8111888111888
12 76.2777777777778
13 74.7719537100068
14 74.0161725067386
15 71.7908496732026
16 70.7731958762887
17 68.7359198998748
18 67.8024691358025
19 65.6937799043062
20 64.6117647058824
21 62.2675736961451
};
\addlegendentry{\tiny0.4 A (1.1 W)}
\addplot [semithick, forestgreen4416044]
table {%
8 76.9602272727273
9 77.5107296137339
10 78.1442307692308
11 77.7703349282297
12 77.4
13 77.1794871794872
14 76.3815789473684
15 75.25
16 74.1514598540146
17 73.5475113122172
18 71.6666666666667
19 70.7003044802088
20 69.4615384615385
21 67.8947368421052
};
\addlegendentry{\tiny0.6 A (1.63 W)}
\addplot [semithick, crimson2143940]
table {%
8 75.0561797752809
9 76.6164874551971
10 77.1696750902527
11 77.4212241941326
12 77.4492753623188
13 76.8368080517613
14 76.7264895908112
15 75.8014184397163
16 75.0561797752809
17 74.4030630003481
18 73.3058984910837
19 72.5840407470289
20 71.7315436241611
21 70.6878306878307
};
\addlegendentry{\tiny0.8 A (2.14 W)}
\end{axis}

\end{tikzpicture}
    \caption{Efficiency measurement of the \gls{pru}, including losses in the rectifier, voltage protection circuit, buck converter, shunt resistor, and load switch.}
    \label{fig:efficiency-pru}
\end{figure*}
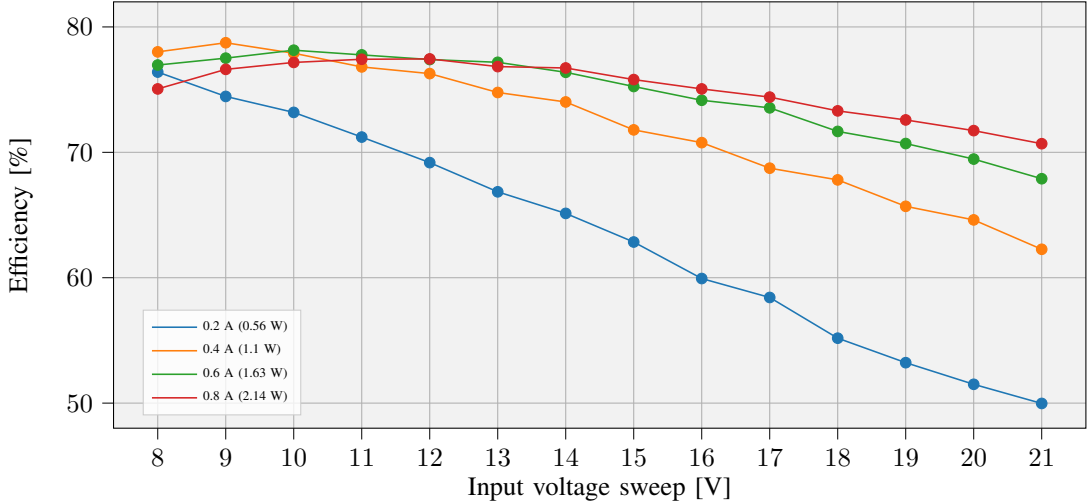

The efficiency of the building blocks from \cref{fig:receiver-charger} was assessed by applying a \gls{dc} voltage behind the LC tank. The experiment utilized an Agilent 6632B power supply functioning as a battery emulator, capable of both sourcing and sinking current. \Cref{fig:efficiency-pru} presents the efficiency corresponding to the input voltage across multiple output power levels. This measurement clearly indicates that achieving an efficiency higher than \SI{80}{\percent} remains unattainable. The (\gls{dc}) rectifier losses constitute a significant contribution to the total losses, along with losses in the converter itself. Furthermore, it is observed that the \gls{dc}/\gls{dc} converter achieves its peak efficiency at around \SI{10}{\volt}.

The charging process concludes when the charge current drops below \SI{200}{\milli\ampere} and the voltage exceeds \SI{2.6}{\volt} for a duration of more than $10$ seconds, indicating the charging process has succeeded. Additionally, if the \gls{uav} leaves the node earlier than expected, the \gls{iot} node transitions to power-down mode when it observes that the input voltage drops below \SI{1}{\volt}. 

\subsection{\Acrfull{ptu}}
\label{subsection:ptu-design}

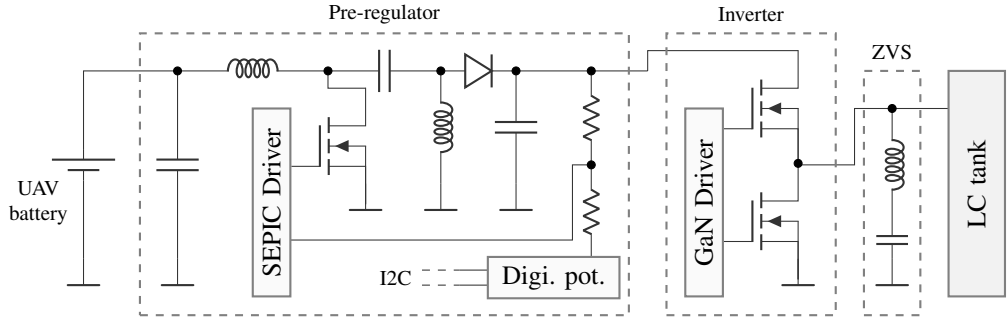
\begin{figure*}[h]
    \centering


\begin{tikzpicture}[american voltages, every text node part/.style={align=center}, /tikz/circuitikz/tripoles/nigfetd/height=.8, /tikz/circuitikz/tripoles/nigfetd/width=.5]

\ctikzset{bipoles/resistor/height=0.15}
\ctikzset{bipoles/resistor/width=0.4}
\ctikzset{bipoles/capacitor/height=0.4}
\ctikzset{bipoles/capacitor/width=0.1}
\ctikzset{bipoles/americaninductor/height=0.35}
\ctikzset{bipoles/americaninductor/width=0.8}

\ctikzset{resistor=american}
\tikzset{block/.style = {rectangle, draw=black!50, fill=black!5, thick, minimum width=3cm, minimum height = 0.75cm}}
\tikzset{blocksmall/.style = {rectangle, draw=black!50, fill=black!2, thick, minimum width=2.5cm, minimum height=0.5cm}}
\tikzset{digipot/.style = {rectangle, draw=black!50, fill=black!2, thick, minimum width=1.75cm, minimum height=0.5cm}}

\tikzset{prereg/.style = {rectangle, dashed, draw=black!50, thick, minimum width=6.5cm, minimum height=3.75cm}}
\tikzset{inverter/.style = {rectangle, dashed, draw=black!50, thick, minimum width=2.25cm, minimum height=3.75cm}}
\tikzset{zvs/.style = {rectangle, dashed, draw=black!50, thick, minimum width=0.75cm, minimum height=3.25cm}}

\tikzset{fontscale/.style = {font=\relsize{#1}}}

\draw[color=black!80]

(-1,2.5) -- (-2.25,2.5) to [battery1] (-2.25,0)
(-2.25,0) node[rground]{}

(-1,0) to [C, /tikz/circuitikz/bipoles/length=40pt](-1,2.5)
(-1,0) node[rground]{}
(-1,2.5) to [L, /tikz/circuitikz/bipoles/length=30pt](1,2.5)
(1.5,1.5) node[nigfete,scale=1] (nmos1) {} 
(1.5,1) node[rground]{}
(nmos1.D) -| (1,2.5)
(1,2.5) to [C, /tikz/circuitikz/bipoles/length=40pt](2.5,2.5)
(2.5,2.5) to [L, /tikz/circuitikz/bipoles/length=30pt](2.5,1)
(2.5,1) node[rground]{}
(2.5,2.5) to [D, /tikz/circuitikz/bipoles/length=25pt](3.5,2.5)
(5.25,2.5) -- (3.5,2.5) to [C, /tikz/circuitikz/bipoles/length=40pt](3.5,1)
(3.5,1) node[rground]{}
(4.5,2.5) to [R, /tikz/circuitikz/bipoles/length=40pt](4.5,1.25) to [R, /tikz/circuitikz/bipoles/length=40pt] (4.5,0)
(4.5,1.25) -| (4.25,1.25) |- (0.5,0.25)

(7.25,0.5) node[nigfete,scale=1] (nmos2) {} 
(7.25,2) node[nigfete,scale=1] (nmos3) {} 
(5.25,2.5) |- (nmos3.D)
(7.25,0) node[rground]{}

(7.25,1.25) -- (8,1.25) |- (9.25,2)
(8.5,2) to [L, /tikz/circuitikz/bipoles/length=30pt](8.5,0.5) to [C, /tikz/circuitikz/bipoles/length=30pt] (8.5,0)
(8.5,0) node[rground]{}

(2.75,-0.15) -- (3.125,-0.15)
(2.75,-0.35) -- (3.125,-0.35)
;

\draw[color=black!80, dashed]
(2.25,-0.15) -- (2.75,-0.15)
(2.25,-0.35) -- (2.75,-0.35)
;

\filldraw[black] (-1,2.5) circle (1.5pt) node[anchor=west]{};
\filldraw[black] (1,2.5) circle (1.5pt) node[anchor=west]{};
\filldraw[black] (2.5,2.5) circle (1.5pt) node[anchor=west]{};
\filldraw[black] (3.5,2.5) circle (1.5pt) node[anchor=west]{};
\filldraw[black] (4.5,2.5) circle (1.5pt) node[anchor=west]{};
\filldraw[black] (4.5,1.25) circle (1.5pt) node[anchor=west]{};
\filldraw[black] (8.5,2) circle (1.5pt) node[anchor=west]{};
\filldraw[black] (nmos2.D) circle (1.5pt) node[anchor=west]{};

\node [blocksmall, rotate=90] at (0.25,0.75) () {SEPIC Driver};
\node [blocksmall, rotate=90] at (6,0.75) () {GaN Driver};
\node [digipot, rotate=0] at (4,-0.25) () {Digi. pot.};
\node [block, rotate=90] at (9.625,1) () {LC tank};

\node [prereg] at (1.75,1.125) () {};
\node [inverter] at (6.625,1.125) () {};
\node [zvs] at (8.5,0.875) () {};

\node [scale=0.8, rotate=0] at (-2.85,0.75) () {UAV\\ battery};
\node [scale=0.8, rotate=0] at (1.9,-0.25) () {I2C};

\node [scale=0.8, rotate=0] at (1.75,3.25) () {Pre-regulator};
\node [scale=0.8, rotate=0] at (6.625,3.25) () {Inverter};
\node [scale=0.8, rotate=0] at (8.5,2.75) () {ZVS};

\end{tikzpicture}%

    \caption{\Gls{ptu} hardware design and components. This board contains additional hardware to measure the voltage and current levels before and after the pre-regulator.}
    \label{fig:transmitter}
\end{figure*}

The central function of the \gls{ptu} board is to generate a \SI{6.78}{\mega\hertz} sine wave with variable amplitude. Both pre-regulator and inverter board from \cref{fig:uav-iot-full-render} needs to be as compact and lightweight as possible. The main functions are discussed in this section. The \gls{ptu} design, illustrating its main components, is depicted in \cref{fig:transmitter}.

The pre-regulator board from \cref{fig:uav-iot-full-render} gets its voltage from the \gls{uav} battery and forwards this voltage to the flight controller. Moreover, the battery voltage is on-board wired to a \gls{sepic} buck-boost converter, which serves as the pre-regulator to modify the input voltage of the half-bridge inverter to strengthen or weaken the alternating magnetic field. The pre-regulator output voltage is adjustable with a programmable potentiometer. This latter is controlled by the \gls{mcu} via an I2C interface. The pre-regulator input, output current and \gls{uav} battery voltage and current on the power supply board can additionally be monitored through this I2C interface.

\Gls{zvs} or soft switching is ensured by a half-bridge inverter combined with an LC resonant circuit to turn the high and low side \glspl{fet} at zero voltage on and off, meaning that the efficiency will be higher compared to hard switching approaches~\cite{xue2017single}. Since the frequency does not change over time in an \gls{mrc} system, unlike other \gls{ipt} implementations, soft switching can be implemented in a low-complex manner, as the \gls{zvs} components do not need to undergo changes during operation. To further reduce switching losses, \gls{gan} \glspl{fet} were selected. The EPC8010 \glspl{fet} are driven by a \gls{gan} driver.

The presented \gls{ptu} design is partially based on the evaluation board EPC9512 \cite{epc9512}, with several modifications to improve the control possibilities and compactness. These modifications include the addition of an \gls{mcu}, the reduction from a full-bridge inverter to a half-bridge inverter, a more appropriate and down-scaled pre-regulator implementation and a separation of the pre-regulator and the power inverter to efficiently utilize the limited \gls{uav} space. As presented in \cref{fig:uav-iot-full-render}, the location of the supply circuit differs from that of the inverter circuit, resulting in two separated PCB designs connected with each other by a flat cable. 

Similar to the \gls{pru}, an NRF52832 NORDIC \gls{sysoc} is selected to transceive \gls{ble} commands with the logger and the \gls{pru}. Additionally, it controls the pre-regulator output voltage, reads current levels, and voltage levels. By combining the current and voltages originating from the \gls{pru}, the \gls{ptu} can calculate the efficiency in real time.

\subsection{Data communication}
\label{subsection:Data_com}

The data communication, presented in \cref{fig:high-level-overview}, consists of the \gls{afhd2a} communication between the remote controller and \gls{uav}, \gls{iot} \gls{lpwan} communication primarily for sensor data, and the \gls{pru}-\gls{ptu}-logger communication. The latter enables real-time link efficiency improvements during the charging process, and is here briefly explained.

During the charging process, the equivalent load of the battery cell $R_{cell}$ will vary over time and both the voltage across and the current through the cell are functions of time. The equivalent \gls{ac}-load $R_L$, after the secondary LC tank, can be adjusted by changing the pre-regulator voltage. With increasing $V_S$, the voltage at the output of the rectifier will increase, causing the current in the DC-DC converter to decrease, thereby altering the load $R_L$ on the secondary side. To facilitate these real-time link efficiency improvements, the \gls{pru} should provide the \gls{ptu} with information about the rectifier voltage, charging current, and charging voltage via the communication link. This enables the \gls{ptu} to adjust the pre-regulator voltage.

This paper primarily focuses on the hardware design and the assessment of achievable efficiency and sustainability aspects. The communication between PRU and PTU can be further expanded in future work. To support interoperability, the Airfuel alliance resonant baseline system specification can be considered~\cite{iec63028}. This specification outlines the interface for coupled \gls{wpt}, including \gls{mrc} systems. 

During the design, validation phase, and also the experimental measurements in \cref{section:measurements}, the communication was limited to sending advertising unidirectional messages from the \gls{ptu} and \gls{pru} to the logger.

\section{System validation and experimental performance assessment}
\label{section:measurements}

The designs and implementations proposed in \cref{section:design} were integrated in the full system and can be consulted in our repository~\cite{designRepo}. The \gls{wpt} system is validated in this section. During the measurements, all components were placed on the UAV to better approximate real-world conditions. 

\begin{figure*}[h]
    \centering
    \includegraphics[width=0.8\textwidth]{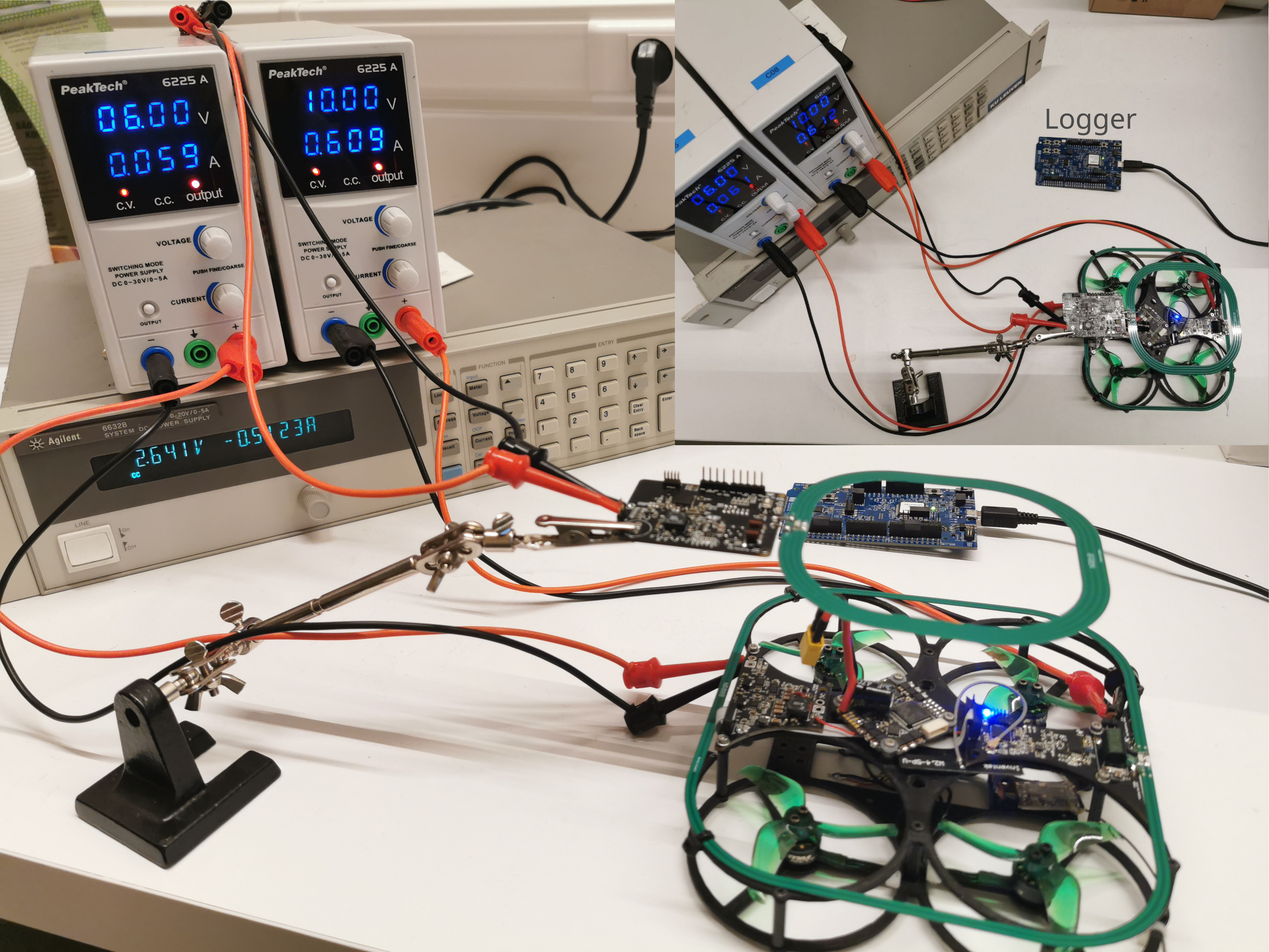}
    \caption{Measurement setup: The left and right voltage sources supply the digital logic and the half-bridge inverter, respectively. The lower power supply acts as the battery emulator.}
    \label{fig:meas-setup}
\end{figure*}

The Agilent 6632B power supply, mentioned in \cref{subsection:pru-design}, is reused to measure the consumed power at the output of the \gls{pru} charger and act as an \gls{lto} cell. The half-bridge inverter in combination with  the transmit coil is crucial for generating the alternating magnetic field. The \gls{sepic} \gls{dc}/\gls{dc} converter, situated on this \gls{ptu}, is not used in the measurements. Alternatively, a voltage source is directly connected to the input of the half-bridge inverter or more specifically to the drain of the high-side \gls{gan} \gls{fet}. This allows us to manually adjust the supply voltage and consequently, adjust the strength of the alternating magnetic field. The measurement setup is depicted in \cref{fig:meas-setup}. 

\begin{figure*}[h]
    \centering
\begin{tikzpicture}

\definecolor{darkgray176}{RGB}{176,176,176}
\definecolor{darkorange25512714}{RGB}{255,127,14}
\definecolor{steelblue31119180}{RGB}{31,119,180}

\begin{axis}[
width=0.8\textwidth,
height=0.4\textwidth,
axis background/.style={fill=gray!10},
tick align=outside,
tick pos=left,
x grid style={darkgray176},
xlabel={Vertical coil distance [mm]},
xmajorgrids,
xmin=47.5, xmax=102.5,
xtick style={color=white},
y grid style={darkgray176},
ylabel=\textcolor{steelblue31119180}{Efficiency [\%]},
ymajorgrids,
ymin=10, ymax=40,
ytick style={color=black}
]
\addplot [draw=steelblue31119180, fill=steelblue31119180, mark=*, only marks]
table{%
x  y
50 39.1759067761457
60 32.4109801263586
70 27.933804332413
80 21.6999110659383
90 17.1362216542939
100 13.2687751107448
};
\addplot [semithick, steelblue31119180]
table {%
50 39.1759067761457
60 32.4109801263586
70 27.933804332413
80 21.6999110659383
90 17.1362216542939
100 13.2687751107448
};
\end{axis}

\begin{axis}[
width=0.8\textwidth,
height=0.4\textwidth,
axis y line=right,
tick align=outside,
x grid style={darkgray176},
xmin=47.5, xmax=102.5,
xtick pos=left,
xtick style={color=black},
y grid style={darkgray176},
ylabel=\textcolor{darkorange25512714}{Inverter input [V]},
ymin=5, ymax=20,
ytick pos=right,
ytick style={color=black},
yticklabel style={anchor=west}
]
\addplot [draw=darkorange25512714, fill=darkorange25512714, mark=*, only marks]
table{%
x  y
50 9.91
60 10.9
70 11.87
80 13.89
90 15.84
100 17.85
};
\addplot [semithick, darkorange25512714]
table {%
50 9.91
60 10.9
70 11.87
80 13.89
90 15.84
100 17.85
};
\end{axis}

\end{tikzpicture}
    \caption{System efficiency with increased inverter input voltage. Measured for a battery voltage of \SI{2.43}{\volt} and charge current of \SI{0.55}{\ampere}, with the pre-regulator bypassed.}
    \label{fig:system-efficiency}
\end{figure*}
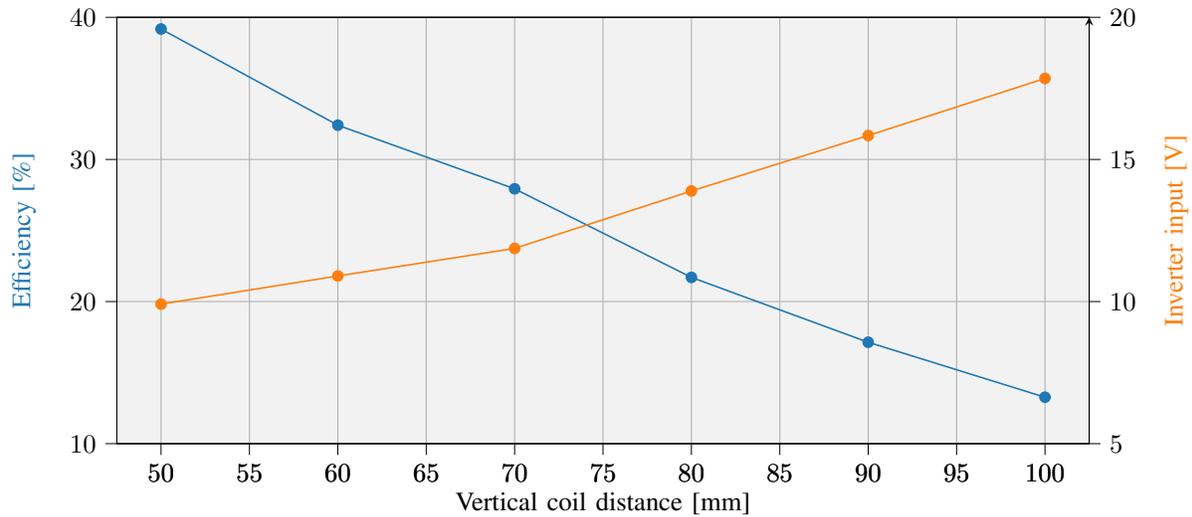

A static measurement has been conducted with the \gls{uav} located on the table. Changing the \gls{pru}-to-\gls{ptu} distance and measuring the input and output power provides us with the measured efficiencies as a function of the vertical distance between the coils. The measurement results are given in \cref{fig:system-efficiency}. 

With an increasing coil-to-coil distance, the voltage after the rectifier will decrease, which can result in an insufficient voltage to power the \gls{pru} charger. From \cref{fig:system-efficiency}, it is evident that $V_S$ must be increased significantly to still transfer the same amount of power with increasing vertical distance between the coils. In these measurements, the optimal point of \SI{10}{\volt} after the rectifier of the PRU was always approached, since this gives the best \gls{pru} efficiency as presented in \cref{fig:efficiency-pru}. The inverter voltage must consistently be elevated with caution. The losses in the \gls{ptu} circuit increase drastically, potentially leading to heating issues. This explains the efficiency drop from \SI{40}{\percent} to less than \SI{10}{\percent} at distances of $50$ and \SI{100}{\milli\meter}, respectively.

In the final system design and operation, the \gls{ble} feedback loop, as depicted in \cref{fig:high-level-overview}, can be engaged to send the charge voltage and current to the \gls{uav}. In conjunction with the pre-regulator voltage, the \gls{uav} may decide to initiate a realignment procedure when the efficiency falls below a certain threshold. Consequently, the feedback loop should guarantee optimal efficiency and prevent reduced charging.

We recall that the current prototype was designed for the scenario where the \gls{uav} hovers during the charging of the \gls{iot} node. It is not suited for the scenario where the \gls{uav} would land to perform the charging, as the coupling factor would become too high, impacting the self-inductance of the coils and thus the resonance frequency of the LC tanks.

\section{Sustainability analysis of the proposed solution}
\label{section:sustainability-analysis}

When examining sustainability, which is a term with a broad meaning, in electronics design conventional methods tend to concentrate solely on the system's energy consumption.
While this may be an important factor in some cases, it may not be in others. In low-power \gls{iot} systems, it has been shown that the environmental impact of the hardware extends far beyond its energy consumption alone~\cite{Ercan2016/08}. In this section, we look beyond the singular metric of energy consumption and comprehensively address the sustainability aspects of the proposed \gls{uav}-based charging solution, embracing a more holistic perspective.

Section~\ref{sec:production_iot_sustainability} addresses the production phase of the devices, while Section~\ref{sec:deployment_sustainability} focuses on the actual deployment and the environmental burden or benefits of the servicing approach.

\subsection{Analysis of overhead in UAV approach}\label{sec:production_iot_sustainability}

We conducted a \gls{lca} to encompass the majority of the solution-related environmental impacts throughout its lifespan. We mainly focus on the electronics at the \gls{iot} node side. The impact of the \gls{uav} has been modeled, excluding the \gls{wpt} circuit. The impact, attributable to the \gls{wpt} circuit, can be distributed over the entire lifespan of the \gls{uav}, given its substantial flight hours. As a result, the \gls{wpt} overhead on the \gls{uav} side is expected to have a negligible contribution to the overall environmental impact.
The modeling is conducted for a cradle-to-gate analysis (production phase) using the Sphera professional and Sphera Extension database XI (Electronics)~\cite{sphera} using the ReCiPe 2016 midpoint (H) method.
As the \gls{eol} processing of electronics has typically a rather small environmental impact compared to the total picture~\cite{carbon-footprint-iot}, and the modeling of these \gls{eol} scenarios is lacking in \gls{lca} databases, we here do not consider the \gls{eol} component. It is important to note that we specifically consider the \gls{gwp} as a metric for sustainability. This only brings a partial analysis and can lead to the so-called `carbon tunnel vision'. There are plenty of other impact categories and metrics that need attention, but are out of scope for this paper.

\begin{figure*}[h!]
\centering

\definecolor{pieblue}{HTML}{8ec6d7}
\definecolor{piegreen1}{HTML}{c8e7a7}
\definecolor{piered}{HTML}{f5949a}
\definecolor{piegreen2}{HTML}{b9cb96}
\definecolor{pieorange}{HTML}{f7ba7a}

\begin{minipage}[b]{0.49\textwidth}
    \centering

    \begin{tikzpicture}
    
    \def\COIL{0.793}
    \def\ICS{0.298}
    \def\PASSIVES{0.347}
    \def\PCB{0.182}
    \def\BATTERY{0.0363}

    \def\SUM{1.66}

    \tikzset{
         lines/.style={draw=none},
    }
    
    \tikzstyle{every node}=[font=\scriptsize]
    
     \pie [/tikz/every pin/.style={align=center},
          every only number node/.style={text=white},
          text=pin,
          scale=0.9,
          explode=0.05,
          radius=2.5,
          sum=\SUM,
          color={pieblue, piegreen1, pieorange, piegreen2, piered},
          style={lines}]
            {\COIL/PCB Coil,
             \ICS/ICs, 
             \PASSIVES/Passives,
             \PCB/PCB,
             \BATTERY/Battery}

\node[] at (0,4.75) {Low-power IoT};
\node[] at (0,4.25) {Total GWP: \SI{\SUM}{kgCO2eq}};
    
    \end{tikzpicture}

\end{minipage}
\hfill
\begin{minipage}[b]{0.49\textwidth}
    \centering

    \begin{tikzpicture}
    
    \def\COIL{0.793}
    \def\ICS{0.298}
    \def\PASSIVES{0.347}
    \def\PCB{0.182}
    \def\BATTERY{0.787}

    \def\SUM{2.41}
    
    \tikzset{
         lines/.style={draw=none},
    }
    
    \tikzstyle{every node}=[font=\scriptsize]
    
     \pie [/tikz/every pin/.style={align=center},
          every only number node/.style={text=white},
          text=pin,
          scale=0.9,
          explode=0.05,
          radius=2.5,
          sum=\SUM,
          color={pieblue, piegreen1, pieorange, piegreen2, piered},
          style={lines}]
            {\COIL/PCB Coil,
             \ICS/ICs, 
             \PASSIVES/Passives,
             \PCB/PCB,
             \BATTERY/Battery}

\node[] at (0,5) {Medium-power IoT};
\node[] at (0,4.5) {Total GWP: \SI{\SUM}{kgCO2eq}};
    
    \end{tikzpicture}

\end{minipage}

\caption{Contribution of different parts of the  \acrshort{iot} node, modified for \gls{uav}-based charging, to the \acrshort{gwp} (in kgCO2eq).}
\label{fig:piechartgwp}
\end{figure*}
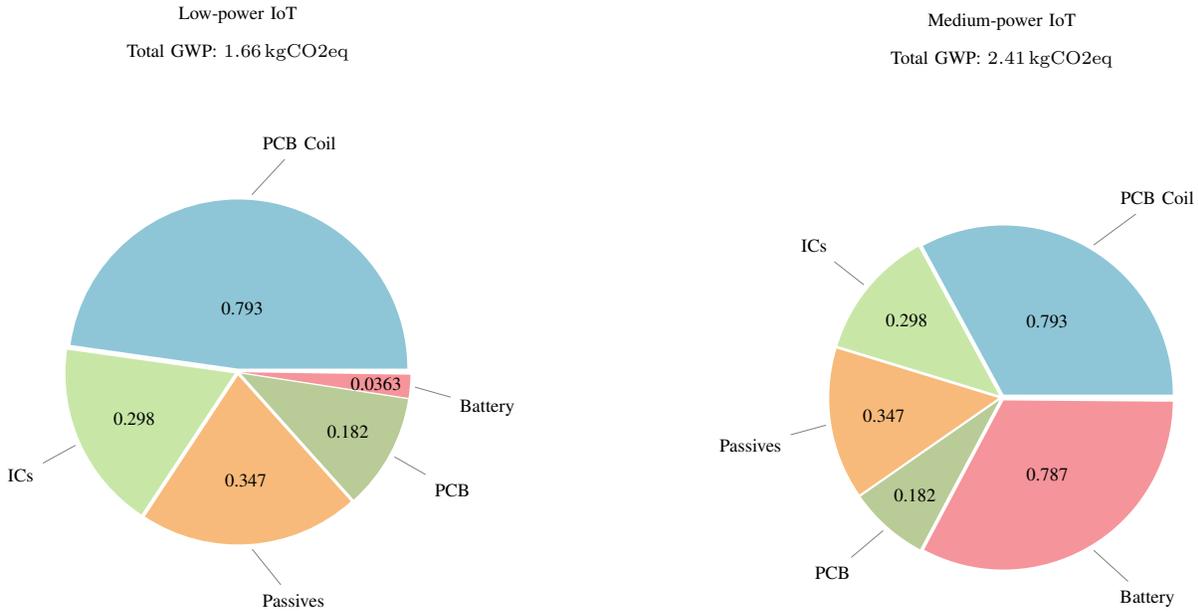

Figure~\ref{fig:piechartgwp} depicts the total overhead needed at the \gls{iot} node side for the \gls{uav}-based servicing approach. 
We here made some assumptions to model the overhead as correctly as possible.
The \gls{ble} \gls{ic} was only taken into account for \SI{50}{\percent} of its \gls{gwp}, since this \gls{ic} can also be used as an application microcontroller.
For the \gls{pcb} modeling, a 2-layer \gls{pcb} with $2/3$ of the original size was used, since the microcontroller area is also partially needed in the application, but the \gls{ble} \gls{pcb} antenna needs to be included specifically for the charging process.
The passive components are modeled according to their closest database match.
Since \gls{lca} models for special types of Lithium-Ion batteries are not broadly available at the time of writing, the \gls{lto} battery is modeled according to a standard Li-Ion battery, recalculated proportional to the weight energy density difference.

Two \gls{iot} systems were considered, one low-power (\SI{10}{\joule\per\day}) and one with medium power consumption (\SI{200}{\joule\per\day}). 
The batteries were chosen accordingly, such that the device could operate for $\pm$ \SI{2}{\months} on a single battery charge. The biggest contribution to the total \gls{gwp} is observed to be in the \gls{pcb}, battery, and passive components, depending on the device's power requirements. 
The low-power system has a \gls{gwp} of \SI{1.66}{kgCO2eq}, from which the \gls{pcb} coil is the biggest contributor. One can observe that the small battery has a very limited impact. In comparison to a standard \gls{iot} device, where the battery plays a substantial role in the device's overall \gls{gwp}, using a compact battery leads to a considerably reduced environmental impact during production. 
Whether this benefit offsets the additional \gls{gwp} from energy harvesting during \gls{uav} deployment is dependent on specific situations.
In the medium-powered system, however, we observe a larger battery capacity is needed to ensure the same lifetime. This translates into an added environmental burden in the manufacturing process.

\subsection{Comparison between UAV and traditional remote IoT deployments}\label{sec:deployment_sustainability}

Examining only the fabrication fails to provide a comprehensive overview. We here look at the total carbon footprint, generated throughout the lifetime of the \gls{iot} node, including potential device or battery recharges/replacements. We compare a typical approach, where an \gls{iot} device would be powered by non-rechargeable batteries, to the novel \gls{uav}-based energy provisioning solution proposed in this work.
Two situations are considered.
\begin{enumerate}
    \item The proposed solution for which the design is detailed in this paper, where the \gls{uav} can recharge an ultra-low-power \gls{iot} device, consuming \SI{10}{\joule\per\day}. This is equivalent to, e.g., a temperature measurement and a \gls{lora} transmission every \SI{30}{\minute}. We here used an \gls{lto} battery with a capacity of \SI{0.144}{\watt\hour} (\SI{60}{\milli\ampere\hour} at \SI{2.4}{\volt})~\cite{LTO_battery_datasheet}.
    \item An extended use-case, with a medium-powered \gls{iot} application, consuming \SI{200}{\joule\per\day}. To be able to power the \gls{iot} device for a sustained period, we here assume a bigger \gls{lto} battery of \SI{3.12}{\watt\hour} (\SI{1.3}{\ampere\hour} at \SI{2.4}{\volt})~\cite{LTO_battery_datasheet}. In this case, the current setup can't charge the battery fast enough while hovering, so a landing alternative is proposed to be able to transfer the full amount of energy.
\end{enumerate}

In the \gls{uav} approach, we consider the base \gls{iot} manufacturing impact together with the additional \gls{wpt} hardware, and the \gls{uav} itself. In the reference case, the same \gls{iot} manufacturing impact is used while also accounting for the impact of the battery.
The \gls{uav} is modeled based on the work in~\cite{carbon-footprint-iot}, assuming optimal usage, i.e., for 400 flight hours before needing replacement, hereby also taking into account battery degradation and thus replacements.
Depending on the goal and size of the \gls{iot} device, the environmental impact ranges from \SI{1}{kgCO2eq} to \SI{10}{kgCO2eq}~\cite{carbon-footprint-iot}. We here assume a relatively simple and small \gls{iot} device with a \gls{gwp} of \SI{3}{kgCO2eq}.
For the non-rechargeable (traditional) \gls{iot}, we considered manual replacement with no environmental impact. This could range from a few grams to almost \SI{1}{kgCO2eq}, when, e.g., driving a car several kilometers. 

The results are depicted in Figure~\ref{fig:sust-uav-analysis}.
In the case of the low-power \gls{iot} node, the \gls{uav}-based servicing/recharging approach is not beneficial compared to a traditional approach (using non-rechargeable batteries) in terms of absolute \gls{gwp}. However, in hard-to-reach applications in, e.g., rural locations, where battery replacement is infrequent, a new device is typically introduced to replace the old one every \SI{5}{\years}. This periodic replacement has a higher ecological impact than the \gls{uav}-based approach starting from the fourth operational year. In a worst-case scenario, where replacements occur annually, the advantages of the \gls{uav}-based approach become more evident and surpass the traditional approach already after \SI{6}{\months}.
In the medium-power \gls{iot}, we observe that the battery becomes significantly large, providing only a \SI{6}{\months} lifespan with a battery of \SI{10}{\watt\hour}. This is a scenario where the \gls{uav}-based approach could bring significant benefit in reducing the total environmental impact. However, it is worth noting that such applications may also opt for rechargeable batteries, as employing very large non-rechargeable batteries becomes impractical.

\usetikzlibrary{positioning}

\begin{figure*}[h!]
\centering

            \definecolor{crimson2143940}{RGB}{214,39,40}
            \definecolor{darkgray176}{RGB}{176,176,176}
            \definecolor{darkorange25512714}{RGB}{255,127,14}
            \definecolor{forestgreen4416044}{RGB}{44,160,44}
            \definecolor{lightgray204}{RGB}{204,204,204}
            \definecolor{steelblue31119180}{RGB}{31,119,180}

\begin{minipage}[b]{0.49\textwidth}

            \begin{tikzpicture}

                \begin{axis}[
                title={Low-power IoT: \SI{10}{\joule\per\day}},
                axis background/.style={fill=gray!10},
                legend to name={mylegend},
                legend cell align={left},
                legend style={ at={(0.5,-0.3)},anchor=north, fill opacity=0.8, draw opacity=1, text opacity=1, draw=white!80.00000!black, /tikz/column 2/.style={column sep=5pt}, /tikz/column 3/.style={column sep=5pt},/tikz/column 4/.style={column sep=5pt}},
                legend style={draw=none, nodes={scale=1, transform shape}},
                legend columns=2, 
                tick align=outside,
                tick pos=left,
                x grid style={darkgray176},
                xlabel={Operational time [Years]},
                xmajorgrids,
                xmin=0, xmax=10,
                xtick style={color=black},
                y grid style={darkgray176},
                ylabel={GWP [kgCO2eq]},
                ymajorgrids,
                ymin=0, ymax=20,
                ytick style={color=black}
                ]
                
                \addplot [draw=steelblue31119180, fill=steelblue31119180, thick] coordinates
                {%
                (0.0,4.696288)
                (15.0,6.819628185185185)
                };
                \addlegendentry{\tiny Recharge by UAV (hovering/landing)}
            
            
                \addplot [draw=darkorange25512714, fill=darkorange25512714, thick] coordinates
                {
                (0.0,3.0)
                (15.0,3.7710625)
                };
                \addlegendentry{\tiny Typical IoT device (battery replacement)}
            
                \addplot [draw=forestgreen4416044, fill=forestgreen4416044, thick] coordinates
                {
                (0.0,3.0)
                (15.0,48.7710625)
                };
                \addlegendentry{\tiny Typical IoT device (full device replacement after 1 year)}
            
                \addplot [draw=crimson2143940, fill=crimson2143940, thick] coordinates
                {
                (0.0,3.0)
                (15.0,12.7710625)
                };
                \addlegendentry{\tiny Typical IoT device (full device replacement after 5 years)}

            
            
            
            

    \coordinate (uav_start) at (axis cs:7,6);
    \coordinate (uav_text) at (axis cs:6,12);
    \draw [black!80] (uav_start) to[out=110,in=180] (uav_text);
    \node[align=left, anchor=west] at (uav_text) {\scriptsize Hovering};

    \coordinate (sign_start) at (axis cs:0.2,0.5);
    \coordinate (sign_text) at (axis cs:0.2,2.5);
    \draw [black!80, stealth-stealth] (sign_start) to[out=90,in=270] (sign_text);
    \node[align=left, anchor=west] at (axis cs:0.5,1.5) {\scriptsize Production};

                \end{axis}
                
            \end{tikzpicture}

\end{minipage}
\hfill
\begin{minipage}[b]{0.49\textwidth}

        \begin{tikzpicture}

            \begin{axis}[
            title={Medium-power IoT: \SI{200}{\joule\per\day}},
            axis background/.style={fill=gray!10},
            legend cell align={left},
            legend style={ at={(0.5,-0.3)},anchor=north, fill opacity=0.8, draw opacity=1, text opacity=1, draw=white!80.00000!black, /tikz/column 2/.style={column sep=5pt}, /tikz/column 3/.style={column sep=5pt},/tikz/column 4/.style={column sep=5pt}},
            legend style={draw=none, nodes={scale=1, transform shape}},
            legend columns=2, 
            tick align=outside,
            tick pos=left,
            x grid style={darkgray176},
            xlabel={Operational time [Years]},
            xmajorgrids,
            xmin=0, xmax=10,
            xtick style={color=black},
            y grid style={darkgray176},
            ylabel={GWP [kgCO2eq]},
            ymajorgrids,
            ymin=0, ymax=20,
            ytick style={color=black}
            ]

            \addplot [draw=steelblue31119180, fill=steelblue31119180, thick] coordinates
            {%
            (0.0,5.44624)
            (15.0,7.130013076923077)
            };

            \addplot [draw=darkorange25512714, fill=darkorange25512714, thick] coordinates
            {
            (0.0,3.0)
            (15.0,18.42125)
            };
        
            \addplot [draw=forestgreen4416044, fill=forestgreen4416044, thick] coordinates
            {
            (0.0,3.0)
            (15.0,63.42124999999999)
            };
        
            \addplot [draw=crimson2143940, fill=crimson2143940, thick] coordinates
            {
            (0.0,3.0)
            (15.0,27.42125)
            };


    \coordinate (uav_start) at (axis cs:5,5.7);
    \coordinate (uav_text) at (axis cs:6.5,3);
    \draw [black!80] (uav_start) to[out=270,in=180] (uav_text);
    \node[align=left, anchor=west] at (uav_text) {\scriptsize Landing};

    \coordinate (sign_start) at (axis cs:0.2,0.5);
    \coordinate (sign_text) at (axis cs:0.2,2.5);
    \draw [black!80, stealth-stealth] (sign_start) to[out=90,in=270] (sign_text);
    \node[align=left, anchor=west] at (axis cs:0.5,1.5) {\scriptsize Production};
    
            \end{axis}
            
        \end{tikzpicture}

\end{minipage}

\ref{mylegend}
    
\caption{Comparison in average environmental impact of low- and medium-powered IoT devices with different servicing methods}
\label{fig:sust-uav-analysis}
\end{figure*}
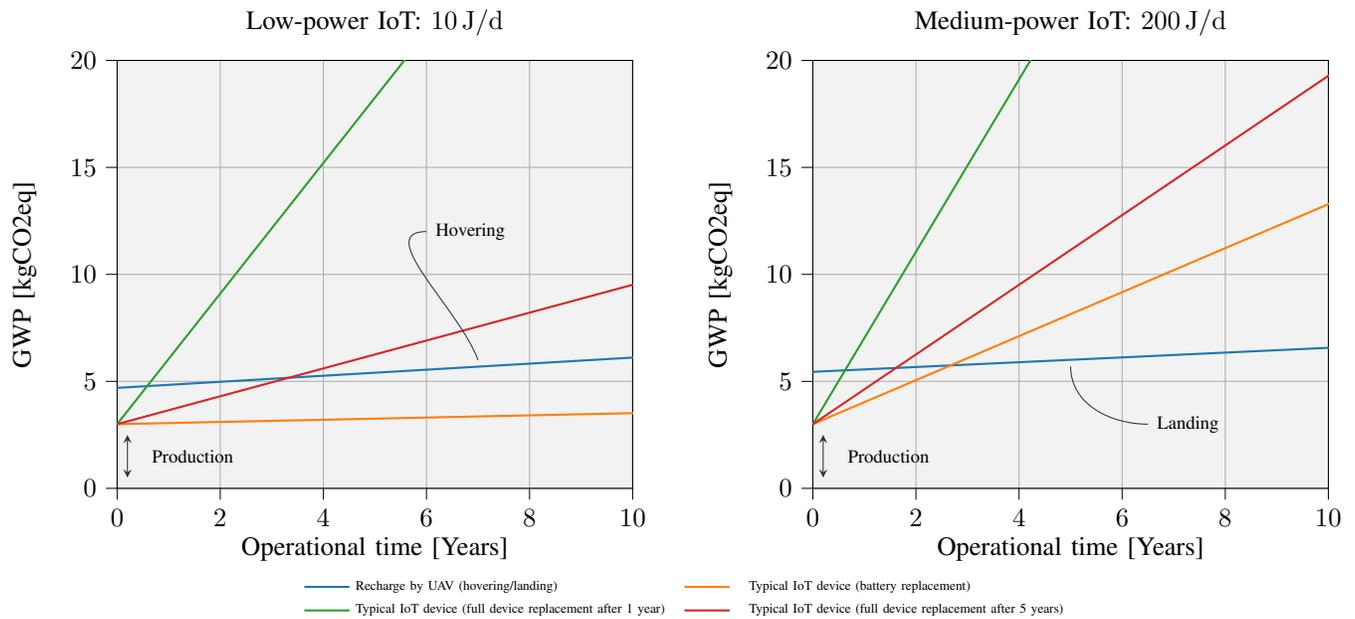

\subsection{Concluding remarks and future opportunities regarding sustainable design}

While the proposed approach may show a higher environmental impact in terms of \gls{gwp} compared to conventional ultra-low-power \gls{iot} systems, its sustainability advantage lies in overcoming the inherent limitations of conventional systems, where the battery size restricts the device's lifespan. 
The proposed approach may prove more sustainable for applications with higher energy requirements, utilizing larger batteries at the \gls{iot} node with enhanced energy transfer capabilities.
The system, with its theoretical capability for an infinite extension of the device's lifespan, presents a substantial benefit by potentially reducing the need for device replacements. The sustainability of the \gls{uav}-based system hinges on the durability of its components, suggesting potential longevity over several decades. Despite an initial burden in the \gls{uav} case due to the necessity for additional \gls{wpt} hardware, this drawback may be outweighed by the extended lifespan of devices.
Lifetimes up to \SI{20}{\years} and longer have been envisioned for \gls{iot} applications, for which foreseeing all the energy in batteries from day 1 increasingly becomes very costly, not practical at all due to the consequentially huge battery and comes with a high ecological impact.
Besides environmental considerations, the convenience of fully automated energy provisioning solutions introduces numerous possibilities for future applications.

\section{Conclusion and future work}
\label{section:conclusion}

This work addressed the challenges of ensuring sustained power autonomy for battery-powered remote \gls{iot} devices through a novel and efficient recharging paradigm: \gls{uav}-based charging using a magnetic resonance coupling approach. An analytical study clarified the trade-offs to be made to reach a good efficiency, quantified the expected impact of misalignment between the coils during the charging, and allowed to optimize the parameters of the wireless power transfer components. An actual design compatible with the constraints imposed by the \gls{uav} on which it has to be mounted, was elaborated, implemented, and measured. The \gls{wpt} was validated, demonstrating its feasibility with good transfer efficiency and highlighting its positive implications for longevity and sustainability.
The paper further presented an analysis of the sustainability aspects of the \gls{uav}-based solution, considering not only energy consumption but also taking into account the environmental impact of the hardware itself.
From this we can conclude that the novel approach can not only bring about the longevity of remote \gls{iot} nodes, it can moreover accomplish this at a relatively low ecological footprint, in particular for devices with medium to larger power needs.

In the future, we envision deploying such systems, for example, to support sound monitoring in scenarios where typical computation-intensive tasks are much more demanding, resulting in significantly lower autonomy if a non rechargeable battery cell would have gained preference. We of course intend to extend the proof of concept to an actual flying and hovering \gls{uav}. Further extension can consider higher power densities, and additional options to improve efficiency as suggested in the course of this paper. This involves enhancing the algorithm to adjust the voltage of the pre-regulator so that optimal load conditions are consistently met. We can further investigate whether angular differences between the coils in windy conditions do not significantly impact the charging process. Additionally, the \gls{uav} will need to be equipped with a fine-grained localization system to achieve alignment with the receiver coil. Additionally, an actual measurement campaign can provide clarification on the feasibility and deployability of this \gls{uav}-based charging solution.



\section*{Conflict of Interest Statement}
The authors declare that the research was conducted in the absence of any commercial or financial relationships that could be construed as a potential conflict of interest.

\section*{Author Contributions}

JVM: Writing – original draft, Conceptualization, Formal analysis, Software, Investigation, Methodology, Validation, Visualization SB: Writing – original draft, Formal analysis, Investigation, Methodology JC; Writing – original draft, Conceptualization, Formal analysis, Investigation, Methodology, Visualization LVdP: Writing – review \& editing, Supervision LDS: Writing – review \& editing, Supervision

\section*{Funding}
This  research  was  partially funded by  Flanders  Innovation and  Entrepreneurship  (VLAIO), grant number HBC.2021.0797 and Cochlear Technology Centre. The results are partly funded by the Flanders Innovation \& Entrepreneurship (VLAIO) project E-CONSTRUCT under grant number HBC.2021.0911.

\section*{Acknowledgments}

We would like to express our gratitude to the advisors Peter Cox (Engineer at BelGaN Belgium) and Jan Cappelle (Professor at KU Leuven) for their valuable input, which contributed to the improvement of this work.

\printbibliography

\end{document}